\documentclass[lettersize,journal]{IEEEtran}
\usepackage{amsmath,amsfonts}
\usepackage{algorithmic}
\usepackage{algorithm}
\usepackage{array}
\usepackage{textcomp}
\usepackage{stfloats}
\usepackage{url}
\usepackage{verbatim}
\usepackage{graphicx}
\usepackage{cite}
\usepackage{indentfirst}
\usepackage{multirow}
\usepackage{graphicx}
\usepackage{color}
\usepackage{xcolor}
\usepackage{colortbl}
\usepackage{listings}
\usepackage{enumerate}
\usepackage{amsmath,amsthm,amscd,amssymb,latexsym,upref,stmaryrd}
\usepackage{amsfonts,epsfig}
\usepackage{CJK}
\usepackage{float}
\usepackage{cite}
\usepackage{lipsum}
\usepackage{multicol}
\usepackage{setspace}
\usepackage{bm}
\usepackage{arydshln}
\usepackage{stfloats}
\usepackage{booktabs}
\usepackage{threeparttable}
\usepackage{stackengine}
\newtheorem{remark}{Remark}
\usepackage[utf8]{inputenc}
\usepackage{fancyhdr}

\usepackage[colorlinks,linkcolor=blue,anchorcolor=blue,citecolor=blue]{hyperref}
\usepackage{makecell}

\newcommand{\calosymbol}[1]{\text{\usefont{U}{BOONDOX-calo}{m}{n}#1}}
\newcommand{\xrowht}[2][0]{\addstackgap[.5\dimexpr#2\relax]{\vphantom{#1}}}

\usepackage{subcaption}

\begin{document}

\title{Fourier Transform-based Wavenumber Domain \\ 3D Imaging in RIS-aided Communication Systems}

\author{Yixuan~Huang,~\IEEEmembership{Graduate Student Member,~IEEE,} Jie~Yang,~\IEEEmembership{Member,~IEEE,} Wankai~Tang,~\IEEEmembership{Member,~IEEE,}\\ \  Chao-Kai~Wen,~\IEEEmembership{Fellow,~IEEE,} and Shi~Jin,~\IEEEmembership{Fellow,~IEEE}
\thanks{
Yixuan~Huang is with the National Mobile Communications Research Laboratory, Southeast University, Nanjing 210096, China (e-mail: huangyx@seu.edu.cn).

Jie~Yang is with the Key Laboratory of Measurement and Control of Complex Systems of Engineering, Ministry of Education, and the Frontiers Science Center for Mobile Information Communication and Security, Southeast University, Nanjing 210096, China (e-mail: yangjie@seu.edu.cn).

Wankai~Tang and Shi~Jin are with the National Mobile Communications Research Laboratory and the Frontiers Science Center for Mobile Information Communication and Security, Southeast University, Nanjing 210096, China (e-mail: \{tangwk; jinshi\}@seu.edu.cn).

Chao-Kai~Wen is with the Institute of Communications Engineering, National Sun Yat-sen University, Kaohsiung 80424, Taiwan. (e-mail: chaokai.wen@mail.nsysu.edu.tw).
}
}

\maketitle

\begin{abstract}
Radio imaging is rapidly gaining prominence in the design of future communication systems, with the potential to utilize reconfigurable intelligent surfaces (RISs) as imaging apertures. Although the sparsity of targets in three-dimensional (3D) space has led most research to adopt compressed sensing (CS)-based imaging algorithms, these often require substantial computational and memory burdens. Drawing inspiration from conventional Fourier transform (FT)-based imaging methods, our research seeks to accelerate radio imaging in RIS-aided communication systems. To begin, we introduce a two-stage wavenumber domain 3D imaging technique: first, we modify RIS phase shifts to recover the equivalent channel response from the user equipment to the RIS array, subsequently employing traditional FT-based wavenumber domain methods to produce target images. We also determine the diffraction resolution limits of the system through {\it k}-space analysis, taking into account factors including system bandwidth, transmission direction, operating frequency, and the angle subtended by the RIS.
Addressing the challenge of limited pilots in communication systems, we unveil an innovative algorithm that merges the strengths of both FT- and CS-based techniques by substituting the expansive sensing matrix with FT-based operators. Our simulation outcomes confirm that our proposed FT-based methods achieve high-quality images while demanding few time, memory, and communication resources.
\end{abstract}


\section{Introduction}

Environment sensing is an emerging function in future wireless communication systems \cite{liu2022integrated,li2019wirelessly,yang2022hybrid}. Among the sensing modalities, radio imaging stands out, illustrating the shapes and scattering properties of the targets in the region of interest (ROI). The radio-based imaging technique is unaffected by light conditions, protects human privacy \cite{mehrotra2022degrees,zhang2012sparse}, and supports applications like monitoring, augmented reality, and environmental reconstruction \cite{imani2020review,huang2023joint,li2023integrated}.
While synthetic aperture radar (SAR) imaging has been widely studied as a type of radio imaging \cite{zhu2016frequency,sheen2001three,lopez20003,zhuge2012three}, its techniques are now being adapted to integrated communication and imaging systems \cite{tong2021joint,huang2024ris}. However, communication antenna arrays are typically undersized for high-resolution imaging.
Recent literature use the antennas of distributed user equipments (UEs) \cite{tong2021joint}, but anisotropic scattering can introduce system modeling errors \cite{ccetin2014sparsity,huang2024ris}.
Recently, reconfigurable intelligent surfaces (RISs) have revolutionized future communications by intelligently customizing the radio propagation environments  \cite{tang2021path,chen2022channel}.
The RIS arrays typically possess large apertures with numerous uniformly placed tunable elements, which are beneficial for high imaging resolutions and potentially support non-line-of-sight (NLOS) imaging \cite{aubry2021reconfigurable}. Moreover, RISs, free of active components, are promising for cost-effective and energy-efficient radio imaging \cite{wang2024reconfigurable}.

Among SAR imaging techniques, Fourier transform (FT) and compressed sensing (CS) dominate \cite{fang2013fast,zhang2012sparse}. Recent RIS-aided imaging research has focused on CS-based algorithms \cite{tong2021joint,hu2022metasketch,huang2024ris}. They alter RIS phase shifts, extract channel responses, and establish linear relationships between these responses and the ROI image. Using two-dimensional (2D) sensing matrices, ROI images are generated iteratively, emphasizing their sparsity in three-dimensional (3D) space \cite{tong2021joint,huang2024ris}.
However, this method requires substantial time and memory resources, making it unapplicable for real-time high-resolution imaging.
Conversely, FT-based algorithms decompose the imaging procedure into a sequence of one-dimensional (1D) operations, which can be rapidly executed via fast Fourier transform (FFT) and its inverse (IFFT), ensuring fast imaging with minimal memory cost \cite{sheen2001three,lopez20003,zhuge2012three,zhu2016frequency}.
However, FT-based imaging in RIS-aided setups is underexplored due to RIS elements' reflecting-only nature \cite{tang2021path}. For dynamic metasurface antenna (DMA)-aided imaging, a solution has emerged \cite{pulido2016application}, suggesting reconfigurations of DMA to capture the electromagnetic field at the surface.
Taking cues from this idea, we obtain the equivalent channel response (ECR) between the UE and the RIS array in the designated RIS-aided system by adjusting RIS phase shifts. As a result, our study transforms the RIS-aided imaging issue into a near-field single-input-multiple-output (SIMO) bistatic imaging problem when using a single UE antenna.
We propose an innovative FT-based two-step imaging approach, incorporating the discrete Fourier transform (DFT) matrix for efficient ECR recovery, followed by the traditional FT-based subimage accumulation algorithm (SAA) \cite{zhu2016frequency} to derive the 3D ROI image.

The proposed FT-based imaging algorithm is designed to retrieve the ROI image with reduced time and memory costs, thanks to the efficient computation of FFT and IFFT. However, it necessitates a significant number of pilots for ECR recovery, which is often unfeasible in real-world communication systems \cite{liu2022integrated}.
To enable the proposed imaging algorithm with inadequate pilots, we propose two foundational methods: the first uses the pseudo-inverse matrix to estimate the ECR, and the second employs block-controlled RIS to mitigate the high demands for pilots.
However, these two methods lead to compromised imaging results when pilots are severely limited.

Alternatively, CS-based approaches directly extract the ROI image from UE-ROI-RIS-access point (AP) channel responses. These methods have been shown to necessitate fewer measurements than the Nyquist criterion requires, leveraging the inherent sparsity of the ROI \cite{tong2021joint,hu2022metasketch,huang2024ris}.
To combine the individual advantages of FT- and CS-based methods, the notion of approximated observation was introduced in \cite{fang2013fast} for SAR imaging.
The sensing matrix in CS-based imaging can be approximated using FT-based operators, thereby eliminating the extensive time and memory requirements of matrix generation and replacing matrix-vector multiplications with efficient FFT and IFFT operations.
Specifically, the time-domain range-Doppler algorithm was applied in \cite{fang2013fast}, and its scope was broadened to wavenumber domain algorithms in \cite{bi2019wavenumber} for SAR imaging systems employing frequency-modulation continuous waves (FMCWs).
However, rare literature has employed this method for RIS-aided imaging with orthogonal frequency-division multiplexing (OFDM) pilots in communication systems, owing to the system discrepancy and waveform mismatch.
In this study, we propose to introduce the solution of approximated observation for RIS-aided imaging issues by integrating our proposed FT-based algorithm with CS-based methodologies, simultaneously reducing the pilot and computation consumptions.

Transforming the imaging issue into a near-field SIMO bistatic imaging problem, we have designed corresponding imaging algorithms.
To evaluate the effectiveness of the proposed algorithms, we employ diffraction resolution limits (DRLs) as our primary metric. DRLs serve as pivotal parameters in imaging systems, informing system design choices like the RIS array deployment, imaging distance, and voxel sizes \cite{fromenteze2021lowering}.
While DRLs have a solid footing in traditional SAR imaging systems \cite{sheen2001three,lopez20003,zhuge2012three}, they tend to focus on far-field monostatic scenarios, which do not fit the near-field bistatic settings in our study.
The widely quoted range DRL in radar and microwave imaging literature, as detailed in \cite{sheen2001three}, is purely inversely proportional to the system bandwidth.
However, recent studies have realized near-field 3D imaging with a single frequency \cite{fromenteze2017single,huang2024ris}, proving that the range DRL in \cite{sheen2001three} does not hold in near fields.
Furthermore, the influences of bistatic imaging on resolutions have been rarely studied.
To address the mismatches between the traditional DRLs and the considered scenario, we propose to derive the DRLs of the RIS-aided system using $k$-space analysis methods \cite{sheen2001three,fromenteze2017single}.
Our findings indicate that the range DRL depends on aspects like the system bandwidth, transmission orientation, operating frequency, and the angular expanse of the RIS array. Conversely, the cross-range DRL largely relies on the wavelength and the RIS-subtended angle.

In summary, our primary contributions include:
\begin{itemize}
\item \textbf{Innovative Two-Step Wavenumber Domain Imaging Algorithm for RIS-Aided Systems:} We transform the RIS-aided imaging problem into a SIMO bistatic imaging scenario by extracting the ECR between the UE and the RIS array using specific RIS phase shifts. Thanks to this transformation, we can utilize the conventional FT-based SAA to create detailed ROI images. Furthermore, RIS-aided NLOS imaging is also studied.

\item \textbf{Effective Imaging with Limited Pilots:} To address the challenge of sparse pilots, we introduce the pseudo-inverse matrix and block-controlled RIS methods. We also compare FT- and CS-based imaging techniques and combine the proposed FT-based algorithm with CS-based methods. Our simulations verify the superior efficiency of our algorithms, characterized by reduced time, memory, and measurement requirements.

\item \textbf{Near-Field SIMO Bistatic DRLs:} We formulated the DRLs for our RIS-aided communication system using $k$-space analysis methods. This sheds light on the interplay between imaging performance and factors such as system bandwidth, transmission orientation, operating frequency, and the angle subtended by the RIS.

\end{itemize}

{\bf Notations}---Scalars, such as $a$, are denoted in italic; vectors, such as $\mathbf{a}$, in bold; and matrices, such as $\mathbf{A}$, in bold capital letters.
The function $\operatorname{vec}(\mathbf{A})$ represents the vectorization of $\mathbf{A}$.
The $\ell_{\bullet}$-norm of $\mathbf{a}$ is denoted by $\|\mathbf{a}\|_{\bullet}$, where ${\bullet}\in\{1,2\}$.
The expression $|a|$ refers to the amplitude of $a$, and $j = \sqrt{-1}$ denotes the imaginary unit.
The notation $\text{diag}(\mathbf{a})$ represents a diagonal matrix with elements from $\mathbf{a}$.
The symbols for transpose and Hermitian operators are given by $(\cdot)^{\rm{T}}$ and $(\cdot)^{\rm{H}}$, respectively. The Hadamard product and division are symbolized by $\odot$ and $\oslash$, respectively.
The notation $\breve{b}(\cdot)$ denotes a continuous function, whereas ${b}(\cdot)$ signifies the discrete version, and $\hat{b}(\cdot)$ presents the estimate of ${b}(\cdot)$.
$\breve{{\cal F}}$, ${\cal F}$, and ${\cal F}^{-1}$ represent the FT, DFT, and inverse DFT (IDFT) operators, respectively.

\begin{figure}
    \centering
    \includegraphics[width=0.77\linewidth]{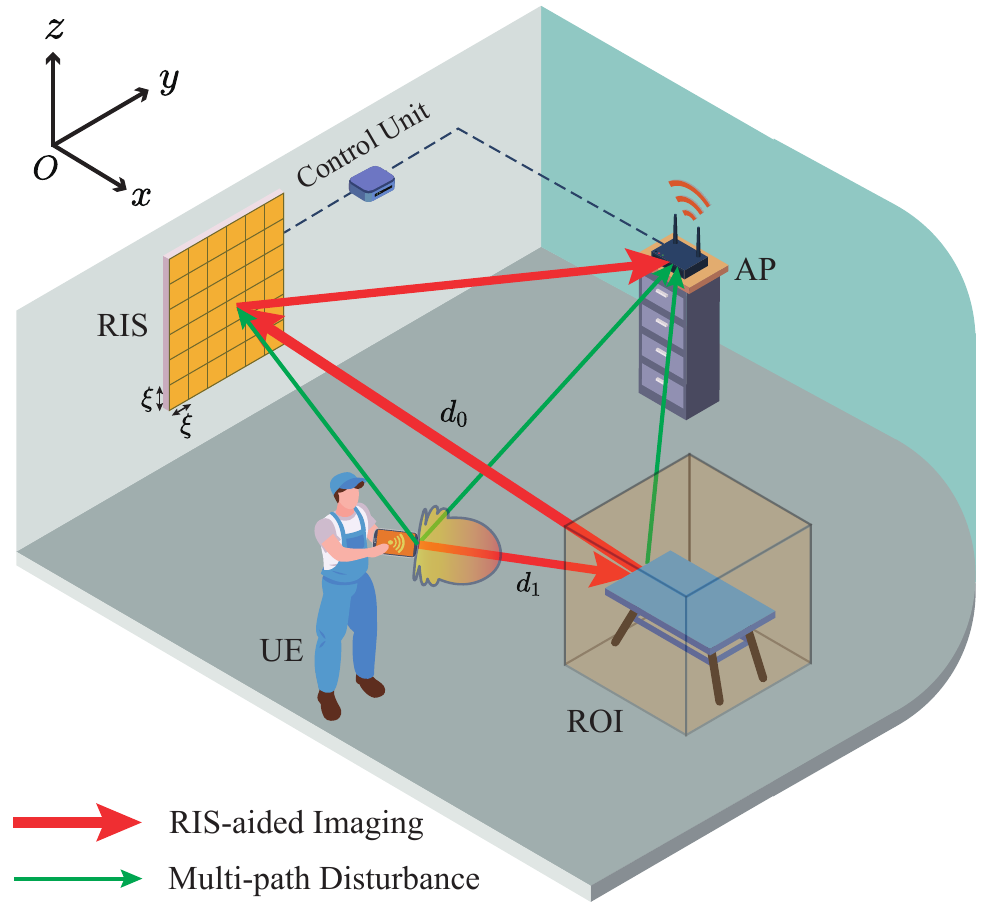}
    \captionsetup{font=footnotesize}
    \caption{Illustration of the considered RIS-aided communication system.}
    \label{fig-environment}
\end{figure}

\section{System Model}
\label{sec-system-model}

We consider a RIS-aided communication system functioning within the 3D space $ [x,y,z]^{\rm{T}} \in \mathbb{R}^3$, as illustrated in Fig. \ref{fig-environment}.
The setup includes a single-antenna UE, an AP, a vertically positioned RIS on the wall, and a target located within the ROI in front of the RIS.
Assuming, for simplicity, that the ROI's center is at the origin, the RIS array lies in the y-z plane, centered at $[-D_0, 0, 0]^{\rm{T}}$.
Given that the ROI is near to the RIS, we infer $D_0 < 2(L_{\max}^{\rm{s}})^2/\lambda_0$. Here, $L_{\max}^{\rm{s}} = \max\{L^{\rm{s}}_{\rm{y}},L^{\rm{s}}_{\rm{z}}\}$ signifies the RIS array's greatest dimension \cite{tang2021path}, and $\lambda_0$ represents the wavelength of the central carrier frequency.
The dimensions $L^{\rm{s}}_{\rm{y}}$ and $L^{\rm{s}}_{\rm{z}}$ represent the RIS array sizes along the y- and z-axis, respectively.
The AP, featuring $A_0$ antennas, coordinates with the RIS, which employs a uniform planar array consisting of $M = M_{\rm{y}} \times M_{\rm{z}}$ elements. Each RIS element measures $\xi\times\xi$ in size.
The phase shifts of the RIS elements are adjusted by the AP through a control unit.
The UE transmits uplink pilot and communication signals \cite{lin20195g}, which are received and processed by the AP to realize joint communication and imaging.
The system bandwidth is set at $B$ with $T_0$ subcarriers, where $T$ subcarriers of them spaced at $\Delta f$ facilitate imaging.
It is presupposed that both the RIS and AP locations are priorly known. Additionally, the UE's location can be precisely discerned using sophisticated UWB and RFID-aided localization methods \cite{huang2024ris}.

In this study, the RIS phase design is time-division multiplexed to support the functions of communication and imaging. During normal communication periods, the RIS phases can be optimized to enhance communication performance, which is a topic widely studied in the literature \cite{tang2021path,chen2022channel}. However, the optimal RIS configuration remains constant in static environments, which may not fully capture the characteristics of the targets. Conversely, during imaging, the RIS phases should vary to capture diverse information about the target, where the corresponding channel state information (CSI) is derived in conjunction with known pilots. When executing the imaging procedure, the UE transmits $R$ pilots on the $T$ subcarriers for imaging, employing $R$ distinct RIS phase shifts across different symbol intervals \cite{chen2022channel}. As a result, the scattering characteristics of the targets can be discerned. \footnote{Conducting imaging may degrade communication performance since the communication resource is partly occupied and the RIS phases are not optimized to maximize communication rates.}
Next, we shift our focus to the imaging process of the system, where the related signal and channel models are presented.

\subsection{Received Signal Model}

Considering that most commercial communication systems employ OFDM signals \cite{liu2022integrated}, we would like to achieve imaging alongside the communication process.
Based on the OFDM signal model, when the pilot $p_{t,r}$ on the $t$-th subcarrier is transmitted at the $r$-th symbol interval, and the $r$-th RIS configuration is employed, the received signals at the AP can be given by
\begin{equation}\label{eq-receive}
\mathbf{y}_{t,r} = \left(\mathbf{h}_{t,r}+\tilde{\mathbf{h}}_{t,r}\right)p_{t,r}+\tilde{\mathbf{n}}_{t,r},
\end{equation}
where $\mathbf{h}_{t,r}$ denotes the channel response of the UE-ROI-RIS-AP path, which is used for imaging.
$\tilde{\mathbf{h}}_{t,r}$ represents the overall channel response of other paths, including the UE-AP path, UE-ROI-AP path, UE-RIS-AP path, and additional multipaths arising from random scatterers and reflectors or those experiencing multiple bounces, as depicted in Fig. \ref{fig-environment}.
$\tilde{\mathbf{n}}_{t,r}$ stands for additive noise.
The pilots are assumed to be demodulation reference signals (DMRS), which are essential for communication data demodulation and could be UE-specific and transmitted on demand \cite{lin20195g,wei20225g}.
Although the design of the pilot sequence may impact channel estimation accuracy \cite{lu2023random,zheng2024waveform,wei2023integrated}, we assume that the CSI has been acquired, and $\mathbf{h}_{t,r}$ has been extracted to assist in imaging,\footnote{
$\mathbf{h}_{t,r}$ can be extracted for imaging by identifying its unique angle of arrival at the AP \cite{huang2021reconf}.
Assuming that the UE antenna is directed towards the ROI during imaging, the energy of the UE-RIS-AP path is minimal, making the extracted RIS-reflected path predominantly $\mathbf{h}_{t,r}$.
When the ROI is located in the NLOS region of the AP, the channel estimation process may be simplified but does not impact the employment of the proposed algorithms.}
as documented in related literature \cite{hu2022metasketch,huang2024ris,tong2021joint}.
Consequently, the extracted UE-ROI-RIS-AP path at the $a$-th AP antenna can be expressed as
\begin{equation}\label{eq-noise}
{s}_{t, r, a} = {h}_{t, r, a} + {n}_{t, r, a},
\end{equation}
where ${h}_{t, r, a}$ is the $a$-th element of $\mathbf{h}_{t, r}$, and
${n}_{t, r, a}$ represents additive noise resulting from channel estimation errors.
The imaging goal is to derive the scattering coefficient image $\breve{\sigma}(x,y,z)$ of the ROI using $\{{s}_{t, r, a}\}$, where $t = 1,2,\ldots, T$, $r = 1, 2, \ldots, R$, and $a = 1,2,\ldots,A_0$.
The value $\breve{\sigma}(x_0,y_0,z_0)$ represents the scattering coefficient of the voxel centered at $[x_0, y_0, z_0]^{\rm{T}}$ with the size of $\delta_x\times \delta_y\times \delta_z$, where $\delta_x$, $\delta_y$, and $\delta_z$ are small values. The scattering coefficient of the voxel equals the root of the energy it scatters when illuminated by a unit energy field \cite{mehrotra2022degrees,broquetas1998spherical,huang2024ris}.

\subsection{UE-ROI-RIS-AP Channel Response Model}
\label{subsec-channel-model}

This subsection elucidates the relationship between ${h}_{t, r, a}$ and the ROI image $\breve{\sigma}(x,y,z)$. It can be expressed as
\begin{equation}\label{eq-min}
h_{t,r,a} =  \sum_{u = 1}^{M_{\rm{y}}} \sum_{v = 1}^{M_{\rm{z}}} g_t {b}_t(y_u, z_v) e^{-j\omega_{r, u, v}} \frac{e^{-jk_td^{\rm{p}}_{u, v,a}}}{(4\pi)^{0.5}d^{\rm{p}}_{u, v, a}},
\end{equation}
where $g_t$ is a constant determined by antenna gains.
The location of the $(u,v)$-th RIS element is $[-D_0, {y}_u, {z}_v]^{\rm{T}}$, whose phase shift under the $r$-th configuration is given by $\omega_{r,u,v}\in[0,2\pi)$.
The term ${b}_t(y_u, z_v)$ denotes the ECR from the UE to the RIS array.
$d^{\rm{p}}_{u, v, a}$ represents the distance from the $(u,v)$-th RIS element to the $a$-th AP antenna, while $k_t=2\pi/\lambda_t$ is the wavenumber for the $t$-th subcarrier.
Denoting $({y}', {z}')$ as the continuous variables corresponding to the discrete RIS element positions $({y}_u, {z}_v)$, the channel response from the UE to the RIS array plane is given as
\begin{equation}\label{eq-b}
\breve{b}_t({y}', {z}') = \iiint \frac{\breve{\sigma}(x,y,z)}{4\pi d_0 d_{1}}  e^{-j k_t (d_0+d_{1}) } dxdydz,
\end{equation}
where $(x,y,z)$ specifies a point's location within the ROI.
The variables $d_1$ and $d_{0}$ represent the distances from the UE to $[x,y,z]^{\rm{T}}$ and from $[x,y,z]^{\rm{T}}$ to $[-D_0, {y}', {z}']^{\rm{T}}$, respectively.
Consequently, the ECR ${b}_t(y_u, z_v)$ is the discrete form of the continuous function $\breve{b}_t({y}', {z}')$, sampled at the $(u, v)$-th RIS element location.
In the next section, \eqref{eq-min} is leveraged to deduce the ECR from the UE to the RIS array. Consequently, \eqref{eq-b} aids in generating the ROI image using the estimated ECR.

\section{FT-based Wavenumber Domain 3D Imaging}
\label{sec-algorithm-two-step}

FT-based imaging algorithms are widely adopted in SAR imaging due to their low computational complexity and superior imaging performance \cite{sheen2001three,lopez20003,zhuge2012three,zhu2016frequency}. However, few studies in the literature have employed FT-based algorithms for RIS-aided imaging. In this section, we introduce an FT-based two-step wavenumber domain imaging algorithm built upon equations \eqref{eq-min} and \eqref{eq-b}. We also determine the DRLs of the RIS-aided system using $k$-space analysis methods.

\subsection{Two-step Imaging Algorithm}
\label{sec-two-step}

In this subsection, we outline the proposed two-step imaging algorithm. The initial step recovers the ECR using specifically designed RIS phase shifts. Subsequently, the second step utilizes wavenumber domain algorithms to construct the image.

\subsubsection{RIS Phase Shift Design and ECR Recovery}
\label{sec-ris-phase-design}

Traditional FT-based wavenumber domain algorithms \cite{sheen2001three,lopez20003,zhuge2012three,zhu2016frequency} require the knowledge of the channel response from the transmitting antenna to the receiving antenna. However, the reflection-only RIS under consideration cannot receive signals. This renders the channel response from the UE to the RIS array unobtainable. In this study, we propose to recover the ECR by varying the RIS phase shifts.

For a compact expression of \eqref{eq-min}, we use $\mathbf{B}_t\in\mathbb{C}^{M_{\rm{y}}\times M_{\rm{z}}}$ to represent the 2D ECR at the $t$-th subcarrier, whose $(u,v)$-th element is ${b}_t(y_u, z_v)$.
Additionally, define $\mathbf{b}_t = {\rm{vec}}(\mathbf{B}_t) \in \mathbb{C}^{M \times 1}$, and let
$\boldsymbol{\omega}_r = [e^{-j\omega_{r,1}}, e^{-j\omega_{r,2}}, \ldots, e^{-j\omega_{r,M}}]^{\rm{T}}$ denotes the $r$-th RIS phase shift configuration.
The free-space channel response from the RIS to the $a$-th AP antenna is denoted by
\begin{equation}\label{eq-h-ris-ap}
\mathbf{h}_{t,a}^{\rm{s,p}} = \left[\frac{1}{(4\pi)^{0.5}d^{\rm{p}}_{m,a}}e^{-jk_td^{\rm{p}}_{m,a}}\right]_{M\times 1},
\end{equation}
where the subscript $m$ represents the $m$-th RIS element and corresponds to the $(u,v)$-th element in \eqref{eq-min}.

In the following discussions, we focus solely on the measurements at the $a^*$-th AP antenna.\footnote{
While multiple AP antennas can offer numerous measurements for ECR recovery, the channels from the RIS array to the antennas are highly correlated due to small antenna spacings. This means that the measurements at each AP antenna offer almost identical information about the ECR, especially when the RIS is in the far field of the AP. Hence, measurements at each AP antenna are processed independently according to Sec. \ref{sec-ris-phase-design} in this study, and their recovery results can be averaged to suppress additive noise.
}
For clarity, we henceforth omit the index $a$ in \eqref{eq-noise}, \eqref{eq-min}, and \eqref{eq-h-ris-ap}.
Using the defined notations, $h_{t,r}$ in \eqref{eq-min} can be represented as $h_{t,r} = g_t \mathbf{b}_t^{\rm{T}} {\rm{diag}}(\boldsymbol{\omega}_r) \mathbf{h}_t^{\rm{s,p}}$. Integrating this expression into \eqref{eq-noise}, we deduce
\begin{equation}\label{eq-min-new}
\begin{aligned}
s_{t,r} & = g_t \mathbf{b}_t^{\rm{T}} {\rm{diag}}(\boldsymbol{\omega}_r) \mathbf{h}_t^{\rm{s,p}} + n_{t,r} \\
& = g_t \boldsymbol{\omega}_r^{\rm{T}} (\mathbf{b}_t \odot \mathbf{h}_t^{\rm{s,p}}) + n_{t,r},
\end{aligned}
\end{equation}
where $\mathbf{h}_t^{\rm{s,p}}$ can be calculated based on known RIS and AP locations. Varying the RIS phase shifts with $R$ distinct configurations, we can consolidate the measurements at the $t$-th subcarrier to derive
\begin{equation}\label{eq-min-new-vec}
\mathbf{s}_{t} = g_t \boldsymbol{\Omega}_R (\mathbf{b}_t \odot \mathbf{h}_t^{\rm{s,p}}) + \mathbf{n}_{t},
\end{equation}
where $\mathbf{s}_{t} = [s_{t,1}, s_{t,2}, \ldots, s_{t,R}]^{\rm{T}}$, $\boldsymbol{\Omega}_R = [\boldsymbol{\omega}_1, \boldsymbol{\omega}_2, \ldots, \boldsymbol{\omega}_R]^{\rm{T}}$, and $\mathbf{n}_{t} = [n_{t,1}, n_{t,2}, \ldots, n_{t,R}]^{\rm{T}}$. Thus, the ECR recovery problem aims to deduce $\mathbf{b}_t$ from $\mathbf{s}_{t}$ with known $\mathbf{h}_t^{\rm{s,p}}$ and $\boldsymbol{\Omega}_R$.

\begin{figure*}
    \centering
    \includegraphics[width=0.85\linewidth]{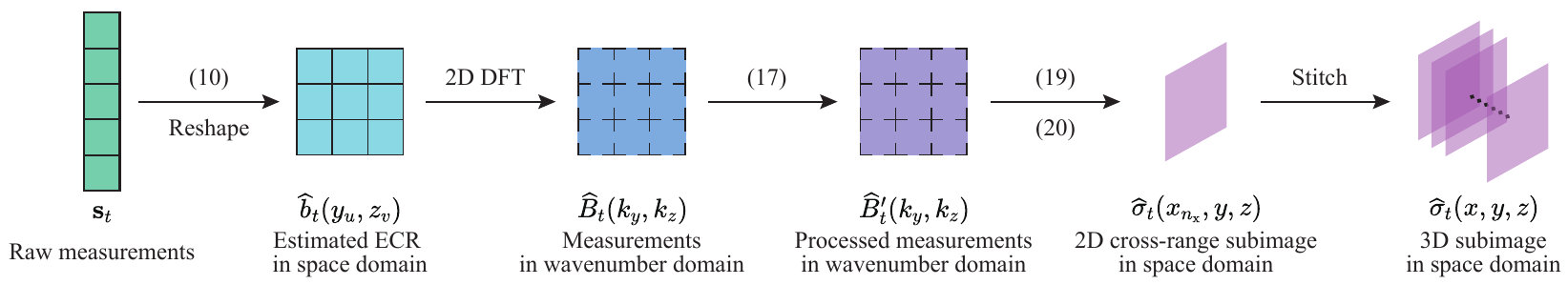}
    \captionsetup{font=footnotesize}
    \caption{Data stream on the $t$-th subcarrier, where $\widehat{b}_t(y_u, z_v)$, $\widehat{B}_t(k_y, k_z)$, $\widehat{B}'_{t}(k_y,k_z)$, $\widehat{\sigma}_t(x_{n_{\rm{x}}},y,z)$, and $\widehat{\sigma}_t(x,y,z)$ represent estimated discrete functions.}
    \label{fig-data-stream}
\end{figure*}

Since $\mathbf{b}_t$ contains $M$ unknown elements, at least $M$ measurements, or a minimum of $M$ pilots with various RIS phase configurations, are needed to retrieve $\mathbf{b}_t$. When $R=M$, assuming that the RIS phase shifts are randomized and $\boldsymbol{\Omega}_R := \boldsymbol{\Omega}_M$ is of full rank, $\mathbf{b}_t$ can be estimated by
\begin{equation}\label{eq-recover-bt-matrix-inverse}
\widehat{\mathbf{b}}_{t} = (\boldsymbol{\Omega}_M^{-1} \mathbf{s}_t) \oslash (g_t \mathbf{h}_t^{\rm{s,p}}).
\end{equation}
However, the matrix inversion in \eqref{eq-recover-bt-matrix-inverse} typically demands a complexity of $O(M^3)$. Considering the large number of RIS elements, $M$, the recovery method in \eqref{eq-recover-bt-matrix-inverse} requires significant computational resources.
Inspired by the intelligent reconfiguration capability of the RIS, $\boldsymbol{\Omega}_M$ can be designed as a DFT matrix, given that the elements in both the DFT matrix and $\boldsymbol{\Omega}_M$ are complex values with unit magnitude.
Furthermore, the condition number of the DFT matrix equals one, achieving robust ECR recovery.
Although $\boldsymbol{\Omega}_M$ can be further optimized to strengthen the signal energy of the UE-ROI-RIS-AP path, we employ the DFT matrix since its inversion can be efficiently implemented with the IFFT algorithm, reducing the computational complexity from $O(M^3)$ to $O(M\log_2M)$.
As a result, \eqref{eq-min-new-vec} can be reframed as
\begin{equation}\label{eq-min-new-vec-22}
\mathbf{s}_{t} = {\cal F}\{ \mathbf{b}_t \odot (g_t\mathbf{h}_t^{\rm{s,p}}) \} + \mathbf{n}_{t}.
\end{equation}
Consequently, ECR recovery can be achieved by
\begin{equation}\label{eq-recover-bt-fft}
\widehat{\mathbf{b}}_{t} = {\cal F}^{-1}\{\mathbf{s}_t\} \oslash (g_t \mathbf{h}_t^{\rm{s,p}}),
\end{equation}
where the IDFT operation can be efficiently implemented with the IFFT algorithm.

By recovering the ECR from the UE to the RIS array, we can equivalently treat the uniformly positioned RIS elements as receiving antennas. Consequently, the initial RIS-aided imaging problem becomes a SIMO bistatic issue, which can be addressed using conventional wavenumber domain algorithms \cite{sheen2001three,lopez20003,zhuge2012three,zhu2016frequency}.
However, potential multiple-input-multiple-output (MIMO)-to-SIMO simplifications and monostatic-to-bistatic phase corrections should be carried out \cite{zhou2019precise}, as traditional radar imaging usually contemplates monostatic MIMO scenarios \cite{sheen2001three,lopez20003,zhuge2012three}.
In this study, we utilize the wavenumber domain SAA introduced in \cite{zhu2016frequency} since it is free from interpolation and focuses on SIMO bistatic imaging situations, thereby simplifying the algorithm's execution.

\subsubsection{Wavenumber Domain SAA}

Next, we detail the SAA \cite{zhu2016frequency} to reconstruct the ROI image from the recovered ECR in \eqref{eq-recover-bt-fft}.
We start by taking the 2D FT of both sides of \eqref{eq-b} with respect to $(y', z')$, yielding
\begin{equation}\label{eq-b-fft}
\breve{B}_{t}(k_y,k_z) = \iiint \frac{\breve{\sigma}(x,y,z)}{4\pi d_1 }  e^{-j k_t d_1} \breve{E}_t(k_y, k_z) dxdydz,
\end{equation}
where $\breve{B}_{t}(k_y,k_z)=\breve{{\cal F}}_{y',z'}\{\breve{b}_t(y', z')\}$.
$\breve{{\cal F}}_{y',z'}$ denotes the 2D FT versus $y'$ and $z'$, whose $k$-space domain variables are $k_y$ and $k_z$, respectively.
The 2D spatial FT of the components tied to $y'$ and $z'$ is given as
\begin{equation}\label{eq-e-t}
\breve{E}_t(k_y, k_z) = \iint \frac{1}{d_{0}}  e^{-j k_t d_{0}} e^{-j y' k_y} e^{-j z' k_z} dy' dz' ,
\end{equation}
with
\begin{equation}\label{eq-dis}
d_{0} = \sqrt{(x + D_0) ^ 2 + (y - y') ^ 2 + (z - z') ^ 2}.
\end{equation}
Applying the method of stationary phase (MSP) \cite{zhuge2012three}, we have
\begin{equation}\label{eq-e-t-2}
\breve{E}_t(k_y, k_z) = \frac{j2\pi}{k_x}e^{-jk_x(x+D_0) - jk_yy - jk_zz},
\end{equation}
where
\begin{equation}\label{eq-kx}
k_x = \sqrt{k_t^2 - k_y^2 - k_z^2}.
\end{equation}
Substituting \eqref{eq-e-t-2} into \eqref{eq-b-fft}, the equation becomes
\begin{equation}
\begin{aligned}\label{eq-b'}
\breve{B}'_{t}(k_y,k_z) & = \iiint \breve{\sigma}'_t(x,y,z)e^{-jk_xx - jk_yy - jk_zz} dxdydz\\
& = \int e^{-j\sqrt{k_t^2 - k_y^2 - k_z^2}x} \breve{{\cal F}}_{y,z}\{\breve{\sigma}'_t(x,y,z)\} dx,
\end{aligned}
\end{equation}
where $\breve{{\cal F}}_{y,z}$ represents 2D FT along the y- and z-axis, and
\begin{align}
\breve{B}'_{t}(k_y,k_z) &= -jk_xe^{jk_xD_0}\breve{B}_{t}(k_y,k_z), \label{eq-b'2} \\[-0.05cm]
\breve{\sigma}'_t(x,y,z) &= \breve{\sigma}(x,y,z) \frac{1}{2d_1} e^{-jk_td_1}. \label{eq-sigma-t}
\end{align}

Referring to \eqref{eq-b'} and employing the SAA in \cite{zhu2016frequency}, we can estimate $\breve{{\sigma}}(x,y,z)$ in the discretized form.
Here, the discrete version of $\breve{{\sigma}}(x,y,z)$ is denoted as ${{\sigma}}(x,y,z)$, and its estimate is represented by $\widehat{\sigma}(x,y,z)$.
The ROI is discretized into $N = N_{\rm{x}} \times N_{\rm{y}} \times N_{\rm{z}}$ voxels.
Using the measurements from the $t$-th single subcarrier, the discrete 3D subimage is reconstructed by recovering a sequence of 2D cross-range subimages, which are derived at $N_{\rm{x}}$ sampled positions along the range direction.
To obtain 2D cross-range subimages along the y- and z-axis, we first estimate the discrete function corresponding to $\breve{\sigma}'_t(x,y,z)$ at the $n_{\rm{x}}$-th sampled position, given as
\begin{equation}\label{eq-sigma-t-estimate}
\widehat{\sigma}'_t(x_{n_{\rm{x}}},y,z) = {\cal F}^{-1}_{k_y,k_z}\left\{\widehat{B}'_{t}(k_y,k_z)e^{j\sqrt{k_t^2 - k_y^2 - k_z^2}x_{n_{\rm{x}}}}\right\},
\end{equation}
where ${\cal F}^{-1}_{k_y,k_z}$ denotes 2D IDFT with respect to the variables $k_y$ and $k_z$.
Then, the 2D cross-range subimage at the $n_{\rm{x}}$-th sampled position can be inferred from \eqref{eq-sigma-t}, given as
\begin{equation}\label{eq-sigma-t-estimate-new}
\widehat{\sigma}_t(x_{n_{\rm{x}}},y,z) = \widehat{\sigma}'_t(x_{n_{\rm{x}}},y,z)\cdot 2d_1 e^{jk_td_1}.
\end{equation}
Stitching the $N_{\rm{x}}$ 2D cross-range subimages along the range direction, we obtain the 3D subimage $\widehat{\sigma}_t(x,y,z)$.
Here, the procedure to obtain $\widehat{B}'_{t}(k_y,k_z)$ is as follows: First, reshape the estimated $\widehat{\mathbf{b}}_{t}$ from \eqref{eq-recover-bt-fft} into matrix $\widehat{\mathbf{B}}_{t}$, deriving $\widehat{b}_t(y_u, z_v)$. Then, compute the 2D DFT of $\widehat{b}_t(y_u, z_v)$ to get $\widehat{B}_t(k_y, k_z)$. Finally, by plugging $\widehat{B}_t(k_y, k_z)$ into \eqref{eq-b'2}, we have $\widehat{B}'_{t}(k_y,k_z)$.
The data stream on the $t$-th subcarrier is illustrated in Fig. \ref{fig-data-stream}.
As \eqref{eq-sigma-t-estimate} holds solely for the $t$-th subcarrier, the subimage obtaining procedure is executed individually for each subcarrier frequency.
By accumulating the subimages at all subcarriers coherently, the final estimate of the ROI image is given as
\begin{equation}\label{eq-final-estimate}
\widehat{\sigma}(x,y,z) = \sum_{t = 1}^T \widehat{\sigma}_t(x,y,z).
\end{equation}
Algorithm \ref{ag1} summarizes the proposed RIS-aided wavenumber domain 3D imaging algorithm. The independence of data processing at each subcarrier allows the proposed algorithm to employ parallel techniques, thereby enhancing imaging speed.

\begin{algorithm}[t]
\caption{RIS-aided two-step wavenumber domain 3D imaging algorithm.}
\label{ag1}
\begin{algorithmic}[1]
        \STATE $\mathbf{input:}$ RIS phase configuration matrix $\boldsymbol{\Omega}_M$, distance $D_0$, and raw measurements $\mathbf{s}_t$, where $t = 1,2,\ldots,T$.

        \STATE  $\mathbf{for} \  t= 1 \ {\rm{to}} \  T \ \mathbf{do}$

        \STATE \hspace{0.5cm} Recover $\widehat{\mathbf{b}}_t$ from $\mathbf{s}_t$ based on \eqref{eq-recover-bt-matrix-inverse} or \eqref{eq-recover-bt-fft}, when $\boldsymbol{\Omega}_M$
        \\ \hspace{1.0cm} is a random or DFT matrix, respectively.

        \STATE \hspace{0.5cm} Reshape $\widehat{\mathbf{b}}_t$ to 2D function $\widehat{b}_t(y_u,z_v)$.

        \STATE \hspace{0.5cm} Perform 2D DFT on $\widehat{b}_t(y_u,z_v)$, deriving $\widehat{B}_{t}(k_y,k_z)$.

        \STATE \hspace{0.5cm} Obtain $\widehat{B}'_{t}(k_y,k_z)$ according to \eqref{eq-b'2}.

        \STATE \hspace{0.5cm} Estimate $\widehat{\sigma}'_t(x_{n_{\rm{x}}},y,z)$ according to \eqref{eq-sigma-t-estimate}.

        \STATE \hspace{0.5cm} Deduce $\widehat{\sigma}_t(x_{n_{\rm{x}}},y,z)$ according to \eqref{eq-sigma-t-estimate-new}.

        \STATE \hspace{0.5cm} Stitch $\widehat{\sigma}_t(x_{n_{\rm{x}}},y,z)$ along x-axis, deriving $\widehat{\sigma}_t(x,y,z)$.

        \STATE $\mathbf{end\ for}$

        \STATE Accumulate the subimages according to \eqref{eq-final-estimate}.

        \STATE $\mathbf{output:}$ the estimated ROI image $\widehat{\sigma}(x,y,z)$.

\end{algorithmic}
\end{algorithm}

\begin{remark}
In the range direction, the voxel size depends on the sampling number $N_{\rm{x}}$ along the x-axis in \eqref{eq-sigma-t-estimate}.
Moreover, the maximum image range is set by \eqref{eq-sampling-frequency} for a given $\Delta f$.
Conversely, in the cross-range direction, the voxel size and the imaged area mirror the RIS element spacing and the RIS array size, respectively, when the numbers of DFT points along the two directions equal the corresponding RIS element numbers.
This is determined by the DFT transform from $\widehat{b}_t(y_u, z_v)$ in the space domain to $\widehat{B}_t(k_y, k_z)$ in the wavenumber domain, and the inverse transform from $\widehat{B}'_{t}(k_y,k_z)$ in the wavenumber domain to $\widehat{\sigma}'_t(x_{n_{\rm{x}}},y,z)$ in the space domain, as illustrated in Fig. \ref{fig-data-stream}.
However, zero-padding around $\widehat{B}_t(k_y, k_z)$ results in oversampling in the space domain, i.e., a reduced voxel size in $\widehat{\sigma}'_t(x_{n_{\rm{x}}},y,z)$.
Similarly, $\widehat{B}_t(k_y, k_z)$ can be oversampled by performing the DFT of zero-padded $\widehat{b}_t(y_u, z_v)$, leading to an enlarged imaged region of the proposed algorithm.
Using IFFT methods, the computational complexity of Algorithm \ref{ag1} is $O(T(M \log_2 M + MN_{\rm{x}}))$ when no zero-padding is used.
\end{remark}

\subsubsection{Analysis about Nyquist Sampling Criterion}

In traditional FT-based algorithms, the Nyquist sampling criterion should be satisfied to achieve high-quality imaging results. This criterion necessitates that the phase shift caused by the path length difference between two adjacent sampling points is less than $\pi\ {\rm{rad}}$ \cite{sheen2001three}.
The maximum spatial and frequency sampling intervals\footnote{
The spatial sampling interval denotes the receiving antenna spacing (i.e., the RIS element spacing $\xi$ in this study), whereas the frequency sampling interval represents the subcarrier spacing $\Delta f$.
} are provided as \cite{zhuge2012three,lopez20003}
\begin{subequations}
\begin{align}
\Delta \bar{d}_{*} = & \lambda_{\min } \frac{\sqrt{{(L^{\rm{s}}_{*}+L^{\rm{i}}_{*})^{2}}/{4}+D_{0}^{2}}}{L^{\rm{s}}_{*}+L^{\rm{i}}_{*}}, \label{eq-sampling-spatial}  \\
\Delta \bar{f} = & \frac{c}{2 L_{\rm{x}}^{\rm{i}}}, \label{eq-sampling-frequency}
\end{align}
\end{subequations}
where $*\in\{{\rm{y}},{\rm{z}}\}$.
$\Delta \bar{d}_{\rm{y}}$ and $\Delta \bar{d}_{\rm{z}}$ denote the maximum sampling intervals along the y- and z-axis, respectively.
$L^{\rm{s}}_{\rm{y}}$ and $L^{\rm{s}}_{\rm{z}}$ are the RIS array sizes along the y- and z-axis, respectively.
$L^{\rm{i}}_{\rm{x}}$, $L^{\rm{i}}_{\rm{y}}$, and $L^{\rm{i}}_{\rm{z}}$ represent the length of the ROI along the x-, y-, and z-axis, respectively.
$\lambda_{\min }$ signifies the minimum wavelength, and $c$ represents the signal propagation speed.
According to \eqref{eq-sampling-spatial}, $\Delta \bar{d}_{}$ exceeds ${\lambda_{\min }}/{2}$, which is approximately ${\lambda_0}/{2}$ due to the narrow bandwidth of communication systems.
As indicated by \cite{tang2021path}, the RIS element size $\xi$ typically falls within the range $[\frac{\lambda_0}{10}, \frac{\lambda_0}{2}]$.
Consequently, the spatial sampling criterion expressed in \eqref{eq-sampling-spatial} can be met by viewing the RIS elements as equivalent receiving antennas.
Moreover, the subcarrier spacing in communication systems is considerably smaller than the center frequency $f_0$ \cite{lin20195g}.
Taking $L^{\rm{i}}_{\rm{x}}=10\lambda_0$ as an example, we find $\Delta \bar{f} = f_0 / 20$.
Thus, the sampling criterion \eqref{eq-sampling-frequency} in the frequency domain can readily be satisfied in our examined scenarios.
Consequently, the imaging problem can be effectively addressed using the proposed two-step imaging algorithm, with the Nyquist sampling criterion being met.

\subsection{DRLs of the Equivalent SIMO Bistatic System}
\label{sec-drl}
This subsection presents the derivation of the DRLs for the equivalent SIMO bistatic system using $k$-space analysis. DRLs measure the smallest distance between two resolvable points in the imaging system, which is solely related to the system configurations but independent of the implemented imaging algorithm and additive noise \cite{zhou2019modern,goodman2005introduction}.
DRLs are crucial for evaluating the performance of the proposed FT-based two-step imaging algorithm, as discussed in Sec. \ref{sec-result-2}, and guide the designs of imaging systems and algorithms.

For simplicity, consider the $\star$-axis, with $\star\in\{x,y,z\}$. Suppose a point target resides at $\star = \star_0$, and its scattering coefficient is given as $\calosymbol{g}(\star) = \delta(\star - \star_0)$, where $\delta(\cdot)$ is the Dirac delta function. The associated wavenumber domain function is
\begin{equation}\label{eq-kspace-fft}
\calosymbol{G}(k_\star) = \breve{{\cal F}}\{\calosymbol{g}(\star)\} = \int_{\star} \delta\left(\star-\star_{0}\right) e^{-j k_{\star} \star} d \star=e^{-j k_{\star} \star_{0}},
\end{equation}
where $\breve{{\cal F}}$ denotes the FT operator.
The objective of wavenumber domain imaging algorithms is to retrieve $\calosymbol{g}(\star)$ from $\calosymbol{G}(k_\star)$ using IFT. However, due to the finite communication bandwidth and RIS array size, the wavenumber domain bandwidth along the $\star$-axis has constraints. Let $k_\star\in[{k_\star^{\rm{min}}}, {k_\star^{\rm{max}}}]$ with $B_{k_\star} = {k_\star^{\rm{max}}} - {k_\star^{\rm{min}}}$, the imaging outcome of the point target, termed the point spread function (PSF), is expressed as
\begin{equation}\label{eq-kspace-ifft}
\begin{aligned}
\hat{\calosymbol{g}}(\star) & = \breve{{\cal F}}^{-1}\{\calosymbol{G}(k_\star)\} = \int_{{k_\star^{\rm{min}}}}^{{k_\star^{\rm{max}}}} e^{j k_{\star} (\star - \star_{0})} d k_\star\\
& = \varepsilon \cdot B_{k_\star} \cdot {\rm{Sa}}\left(\frac{B_{k_\star}}{2}(\star - \star_0)\right),
\end{aligned}
\end{equation}
where $\breve{{\cal F}}^{-1}$ represents IFT. ${\rm{Sa}}(x) = \sin x / x$, and $\varepsilon$ is a complex constant not affecting the zero points of $\hat{\calosymbol{g}}(\star)$. The PSF mirrors a $\rm{Sa}(\cdot)$ function shape, and its first zero point outside the mainlobe is $|\star - \star_0| = 2\pi / B_{k_\star}$. Based on the Rayleigh criterion \cite{zhou2019modern,goodman2005introduction}, the DRL on the $\star$-axis is
\begin{equation}\label{eq-kspace-limit-star}
\delta_\star = \frac{2\pi}{B_{k_\star}}.
\end{equation}

\begin{figure}
    \centering
    \includegraphics[width=0.73\linewidth]{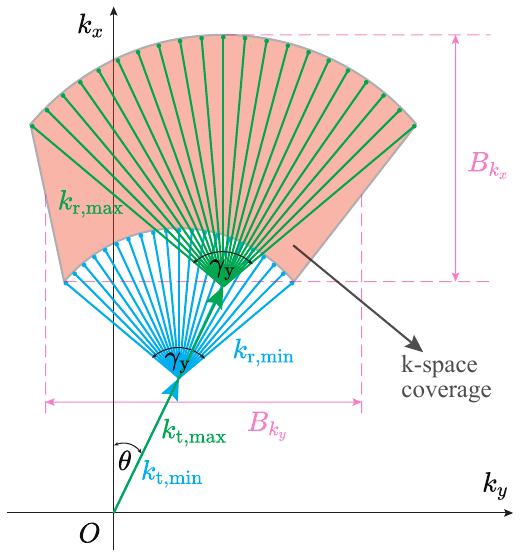}
    \captionsetup{font=footnotesize}
    \caption{Illustration of the $k$-space coverage in the considered system, where $k_{{\rm{t,min}}}$ and $k_{{\rm{r,min}}}$ ($k_{{\rm{t,max}}}$ and $k_{{\rm{r,max}}}$) represent the transmitting $k$-space vector from the UE to the point target and the scattering $k$-space vector from the point target to the RIS array at the minimum (maximum) subcarrier frequency, respectively.}
    \label{fig-kspace}
\end{figure}

We now examine the $k$-space bandwidth by illustrating the $k$-space coverage of the equivalent SIMO bistatic system in Fig. \ref{fig-kspace}, which has been commonly referenced in related literature \cite{sheen2001three,fromenteze2017single}.
For brevity, our focus is on the x-y plane, implying that the findings along the y-axis are analogous to those along the z-axis.
In Fig. \ref{fig-kspace}, the head of a transmitting $k$-space vector (e.g., $k_{\rm{t,min}}$) is linked to the tails of its corresponding receiving vectors (e.g., $k_{\rm{r,min}}$), and the area reached by the heads of the receiving vectors delineates the $k$-space coverage when the Nyquist sampling criterion is satisfied, which is shown as the shaded area.
The length of the k-space vector is proportional to the wavenumber amplitude, and its orientation is contingent upon the signal propagation direction. Owing to the utilization of multiple subcarrier frequencies for imaging, the vector length is variable across different frequencies. With only one antenna at the UE, the transmitting direction is fixed, deviating by an angle $\theta$ from the RIS array's perpendicular bisector. Conversely, the extensive RIS array facilitates a variable receiving direction, spanning an angle $\gamma_{\rm{y}}$ along the y-axis, which is determined by the RIS aperture.

However, the irregular shape of the $k$-space coverage in Fig. \ref{fig-kspace} complicates the calculation in \eqref{eq-kspace-ifft}.
To better detail influential system parameters on the DLRs, we simplify the calculation by approximating the $k$-space coverage as a rectangle \cite{sheen2001three}.
For the range resolution on the x-axis, we maximize $k$-space bandwidth to minimize the DRL, acknowledging the spherical wavefront in proximal fields. For the cross-range DRL on the y-axis, we focus on the center frequency, aligning with \cite{sheen2001three,lopez20003,zhuge2012three}. Hence, the $k$-space bandwidths along the $k_x$- and $k_y$-axis are expressed as
\begin{subequations}
\begin{align}
B_{k_x} &\approx \left(k_{\max}-k_{\min}\right)(1+\cos \theta) + k_{\min}\left(1-\cos \frac{\gamma_{\rm{y}}}{2}\right), \label{eq-kspace-bandwidth-x} \\
B_{k_y} &\approx 2 k_{0} \sin \left(\frac{\gamma_{\rm{y}}}{2}\right), \label{eq-kspace-bandwidth-y}
\end{align}
\end{subequations}
where $k_0$, $k_{\min}$, and $k_{\max}$ are the wavenumbers corresponding to the center, minimum, and maximum frequencies.
To extrapolate the results to 3D space, we replace $\gamma_{\rm{y}}$ in \eqref{eq-kspace-bandwidth-x} with $\gamma_{\max} = \max\{\gamma_{\rm{y}}, \gamma_{\rm{z}}\}$ and redefine $\theta$ in 3D space. The bandwidth on the $k_z$-axis is determined by substituting $\gamma_{\rm{y}}$ with $\gamma_{\rm{z}}$ in \eqref{eq-kspace-bandwidth-y}.
Denoting $f_{\min}$ as the lowest working frequency, the DRLs can be derived based on \eqref{eq-kspace-limit-star}, given as
\begin{subequations}
\begin{align}
\delta_{\rm{x}} & \approx \frac{c}{B(1+\cos \theta)+f_{\min}\left(1-\cos \frac{\gamma_{\max}}{2}\right)}, \label{eq-kspace-limit-x} \\
\delta_{\rm{y}} &\approx \frac{\lambda_{0}}{2 \sin \left(\frac{\gamma_{\rm{y}}}{2}\right)},\quad
\delta_{\rm{z}}  \approx \frac{\lambda_{0}}{2 \sin \left(\frac{\gamma_{\rm{z}}}{2}\right)}. \label{eq-kspace-limit-y}
\end{align}
\end{subequations}

In SIMO bistatic systems, the cross-range DRL is not influenced by the transmitting direction but is contingent on the wavelength $\lambda_0$ and the RIS-subtended angle $\gamma_*$, where $*\in\{{\rm{y}},{\rm{z}}\}$. This angle $\gamma_*$ considers the RIS array size $L_*^{\rm{s}}$ on the $*$-axis and the distance $D_0$ between the RIS array and the ROI. Their relationship is articulated as
\begin{equation}
\sin (\gamma_* / 2)=\sqrt{\frac{(L^{\rm{s}}_* / 2)^{2}}{(L^{\rm{s}}_* / 2)^{2}+D_0^{2}}}=\frac{L^{\rm{s}}_*}{\sqrt{L^{\rm{s}}_*{}^{2}+4 D_0^{2}}},
\end{equation}
when the target is located in front of the RIS array center.
Consequently, larger RIS array sizes or shorter imaging distances enhance the cross-range resolution.

When juxtaposing \eqref{eq-kspace-limit-y} with conventional cross-range DRLs \cite{zhuge2012three,sheen2001three,lopez20003}, it emerges as a SIMO variant of the usual MIMO DRLs. Conversely, the range DRL in \eqref{eq-kspace-limit-x} fundamentally contrasts the traditional limit of $c/2B$ found in far-field monostatic systems \cite{sheen2001three,imani2020review}.
In our approach, the range DRL \eqref{eq-kspace-limit-x} caters to general SIMO bistatic systems, factoring in the signal bandwidth $B$, transmitting direction $\theta$, minimum frequency $f_{\min}$, and angle $\gamma_{\max}$. Increasing $B$, $f_{\min}$, and $\gamma_{\max}$ reduces $\delta_{\rm{x}}$, while elevating $\theta$ does the opposite.
\eqref{eq-kspace-limit-x} further unveils that the range resolution is influenced not just by the bandwidth but also by the operating frequency. Given the typically smaller communication bandwidth $B$ compared to $f_{\min}$, effects of changing $\theta$ can be pronounced for a large $D_0$ (small $\gamma_{\max}$) but minor for a small $D_0$.
It is worth noting that the range DRL in \eqref{eq-kspace-limit-x} often falls below the system imaging capabilities, as the bandwidth in \eqref{eq-kspace-bandwidth-x} sets the maximum range of $k_x$, as shown in Fig. \ref{fig-kspace}. Furthermore, \eqref{eq-kspace-limit-x} reverts to $c/2B$ for monostatic systems when $\gamma_{\max}$ is negligible.

\subsection{Case Study: NLOS Imaging with and without the RIS}
\label{sec-case-study}

\begin{figure}
\centering
\captionsetup{font=footnotesize}
\begin{subfigure}[b]{0.398\linewidth}
\centering
\includegraphics[width=\linewidth]{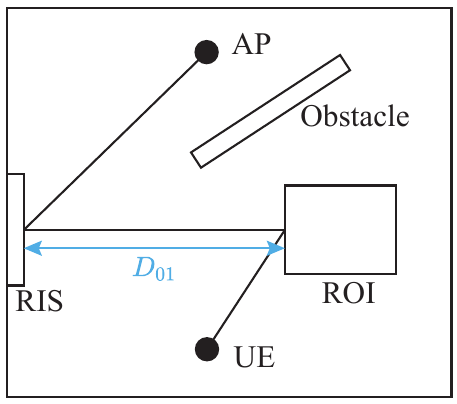}
\caption{Imaging with a RIS}
\label{fig-nlos-scenario-1}
\end{subfigure}
\hfill
\begin{subfigure}[b]{0.58\linewidth}
\centering
\includegraphics[width=\linewidth]{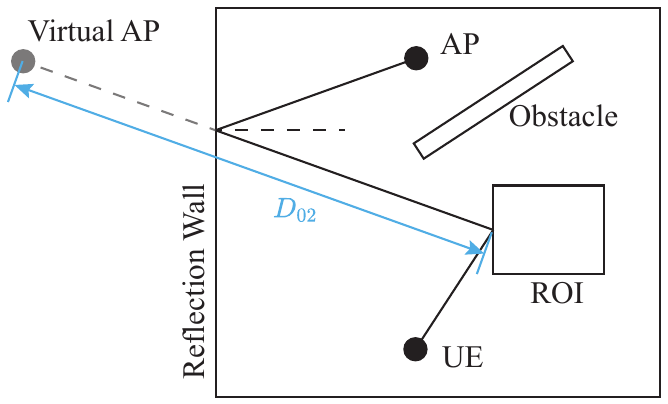}
\caption{Imaging without a RIS}
\label{fig-nlos-scenario-2}
\end{subfigure}

\caption{NLOS imaging scenarios with and without a RIS.}
\label{fig-nlos-scenario}
\end{figure}

The previous sections have examined a general scenario without NLOS paths between the items in Fig. \ref{fig-environment}. However, it should be noted that the proposed RIS-aided imaging algorithm is also effective when the LOS path between the ROI and the AP is obstructed. In this subsection, we demonstrate the advantages of utilizing the RIS for imaging in NLOS conditions, as illustrated in Fig. \ref{fig-nlos-scenario}.

As elucidated in Sec. \ref{sec-two-step}, by utilizing the RIS to assist in imaging, the radio environments can be intelligently adjusted to recover the ECR from the UE to the RIS array. 
Subsequently, the RIS array acts as the imaging aperture with an imaging distance $D_{01}$, as depicted in Fig. \ref{fig-nlos-scenario-1}.
In contrast, in the imaging scenario without the RIS, the signals scattered by the ROI can only be diffused or reflected towards the AP without reconfigurability. 
As a result, the ECR from the UE to the scatterer or reflector may not be recovered, rendering the proposed two-step imaging algorithm unsuitable.
When the RIS is replaced by a scatterer in Fig. \ref{fig-nlos-scenario-1}, employing traditional FT-based algorithms may not facilitate imaging. Alternatively, when signals are reflected by a reflection wall, the ROI can be effectively positioned in the LOS region of the virtual AP, which mirrors the AP with the reflection wall, as shown in Fig. \ref{fig-nlos-scenario-2} and studied in \cite{aladsani2019leveraging}.
Therefore, the traditional SAA can be applied to achieve imaging under this condition, where the antenna array of the AP serves as the imaging aperture, and the imaging distance is $D_{02}$.
Given that $D_{01}$ is significantly less than $D_{02}$, the DRL of the imaging scenario in Fig. \ref{fig-nlos-scenario-1} is expected to be lower than that in Fig. \ref{fig-nlos-scenario-2}, facilitating superior imaging performance.\footnote{According to \eqref{eq-kspace-limit-x} and \eqref{eq-kspace-limit-y}, the DRL is also influenced by the antenna orientations. However, we do not elaborate on this in Fig. \ref{fig-nlos-scenario} and assume that the imaging aperture is orientated towards the ROI for simplicity.}
Moreover, by using the RIS as the imaging aperture, imaging can be accomplished even with a single-antenna AP, thus reducing hardware costs and energy consumption \cite{tang2021path,wang2024reconfigurable}.
According to \cite{huang2023joint,tang2021path}, the received signal energy of the UE-ROI-RIS-AP path in Fig. \ref{fig-nlos-scenario-1} may be lower than that of the UE-ROI-reflector-AP path in Fig. \ref{fig-nlos-scenario-2}.
Nevertheless, this leads to a minor imaging performance decrement when utilizing the proposed Algorithm \ref{ag1}, which is robust against additive noise.

\section{3D Imaging with Limited Pilots}
\label{sec-limited-pilots}

The imaging algorithm introduced in Sec. \ref{sec-algorithm-two-step} necessitates $M$ measurements per subcarrier to deduce the ECR.
However, as the size of the RIS array increases, $M$ becomes significant, leading to high pilot overheads that may be unsuitable for certain communication systems. In this section, we aim to achieve 3D imaging with fewer pilots, thereby minimizing the impacts on communication performance.
Initially, we present two foundational methods: pseudo inverse matrix and block-controlled RIS. Subsequently, we contrast the FT- and CS-based approaches regarding signal models, imaging procedures, and performances. Ultimately, we integrate the FT and CS techniques to facilitate rapid imaging with minimal pilots.

\subsection{Preliminary Methods}
\label{sec-preliminary}

\subsubsection{Method of Pseudo Inverse Matrix}
\label{sec-pseudo-inverse-matrix}

With limited pilots, specifically when $R<M$, $\boldsymbol{\Omega}_R$ is not square, rendering the ECR recovery methods in \eqref{eq-recover-bt-matrix-inverse} and \eqref{eq-recover-bt-fft} inapplicable. A straightforward ECR recovery technique using \eqref{eq-min-new-vec} involves computing the pseudo inverse matrix of $\boldsymbol{\Omega}_R$, which has a computational complexity of $O(M^3)$.
Alternatively, the RIS configuration matrix, $\boldsymbol{\Omega}_R$, can be formulated as the initial $R$ rows of the $M$-dimensional DFT matrix. This means the pseudo inverse matrix of $\boldsymbol{\Omega}_R$ is $\boldsymbol{\Omega}_R^{\rm{H}}$. Consequently, the SAA method can be used for imaging, with an overall computational complexity of $O(T(MR + M\log_2 M + MN{\rm{x}}))$.
However, the approach may not recover the ECR precisely since $\boldsymbol{\Omega}_R^{\rm{H}}\boldsymbol{\Omega}_R\neq \mathbf{I}$ for $R<M$. This discrepancy introduces errors into the estimates of $\mathbf{b}_t$, compromising the image quality.

\begin{figure}
\centering
\captionsetup{font=footnotesize}
\begin{subfigure}[b]{0.28\linewidth}
\centering
\includegraphics[width=\linewidth]{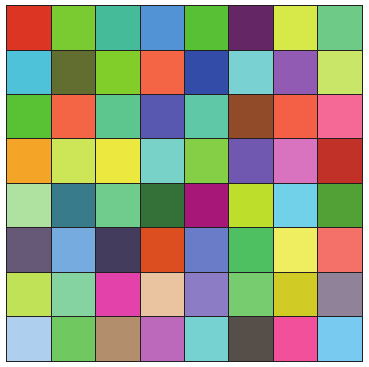}
\caption{$Q=1$}
\label{fig-block-RIS1}
\end{subfigure}
\hfill
\begin{subfigure}[b]{0.28\linewidth}
\centering
\includegraphics[width=\linewidth]{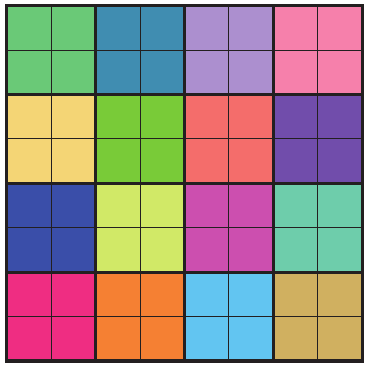}
\caption{$Q=4$}
\label{fig-block-RIS2}
\end{subfigure}
\hfill
\begin{subfigure}[b]{0.28\linewidth}
\centering
\includegraphics[width=\linewidth]{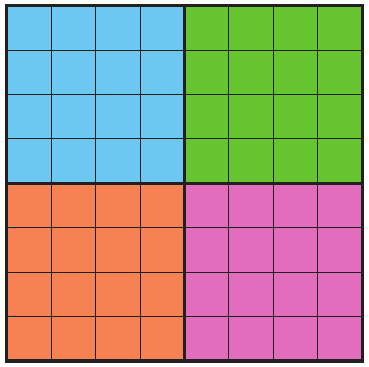}
\caption{$Q=16$}
\label{fig-block-RIS3}
\end{subfigure}

\caption{Illustration of the block-controlled RIS phase shift configuration, where each small square represents a RIS element, and various colors denote different phase shifts.}
\label{fig-block-RIS}
\end{figure}

\subsubsection{Method of Block-controlled RIS}
\label{sec-block-controlled-ris}

The substantial pilot overheads in Algorithm \ref{ag1} stem from the vast number of RIS elements, $M$. In this approach, we suggest adjusting the phase shifts of consecutive $Q$ RIS elements uniformly and treating these $Q$ RIS elements as a single RIS block, as shown in Fig. \ref{fig-block-RIS}. Assuming $M$ is divisible by $Q$ for simplicity, a RIS block's location is deemed the midpoint of its constituent $Q$ RIS elements. This reduces the effective number of receiving antennas (or RIS blocks) to $M_Q = M/Q$, necessitating only $M_Q$ measurements to retrieve the ECR between the UE and the RIS array.
The RIS blocks' phase shifts can be configured as an $M_Q$-dimensional DFT matrix.\footnote{
In a practical scenario, phase shifts within a block might be individually adjusted to offset phase shifts due to varying path lengths from each RIS element to the AP antenna. For instance, if $\omega_{m_Q}$ signifies the intended phase shift for the $m_Q$-th RIS block and $\phi_q$ represents the phase discrepancy between the paths from the $q$-th element of the $m_Q$-th block to the AP and from the block's center to the AP, then the $q$-th element's phase shift in the $m_Q$-th block would be set as $\omega_{m_Q} - \phi_q$. As a result, the ECR for the $m_Q$-th block is an average of the ECRs from its comprising $Q$ elements.
}
Consequently, the efficient ECR recovery approach \eqref{eq-recover-bt-fft} that uses IFFT is applicable.
With this block-controlled RIS strategy, Algorithm \ref{ag1} can be employed for imaging, but the dimensions for both $\mathbf{s}_t$ and the RIS configuration matrix $\boldsymbol{\Omega}_{R}$ shrink to $M_Q$. Hence, the computational load drops to $O(T(M_Q \log_2 M_Q + M_QN_{\rm{x}}))$. Notably, when $Q=1$, Algorithm \ref{ag1} becomes a specific instance of this method. However, since the RIS block's spacing, $\xi_{Q}$, is $Q$ times that of an individual RIS element, equating to $\xi_{Q} = Q\xi$, the spatial sampling criteria in \eqref{eq-sampling-spatial} might be unmet, potentially impairing imaging performance.

\begin{remark}
Both proposed methods face challenges in accurately reconstructing ROI images with limited measurements, but the root causes differ. The first method, as discussed in Sec. \ref{sec-pseudo-inverse-matrix}, suffers due to imprecise ECR recovery, which subsequently degrades imaging results. On the other hand, the second method from Sec. \ref{sec-block-controlled-ris} falters because it does not meet the spatial sampling criterion specified in \eqref{eq-sampling-spatial}. Simulations in Sec. \ref{sec-result-5} indicate that while the first method yields distorted images when $R$ is significantly less than $M$, the second introduces unwanted sidelobes and artifacts.
\end{remark}

\subsection{Comparison between FT- and CS-based Imaging Methods}
\label{sec-ft-cs-compare}

CS theory allows for the reconstruction of sparse signals with fewer measurements than what the Nyquist criterion suggests \cite{zhang2012sparse}. The CS-based imaging technique is explored in \cite{tong2021joint,hu2022metasketch,huang2024ris}, where the ROI image is derived directly from the measurements $\mathbf{s}_t$ instead of the ECR in Algorithm \ref{ag1}.
In this method, the sensing matrix stems from channel modeling, and the image of the ROI is represented by an $N$-dimensional vector $\boldsymbol{\sigma} = [\sigma_1, \sigma_2, \ldots, \sigma_N]^{\rm{T}}$, which is the vector form of the discrete function ${\sigma}(x,y,z)$.
Given the target's sparsity in the ROI, $\boldsymbol{\sigma}$ is inherently sparse. Based on \eqref{eq-b} and \eqref{eq-min-new-vec}, the measurements of the channel response $h_{t,r}$ at the $t$-th subcarrier can be expressed as
\begin{equation}\label{eq-cs-model-single}
\mathbf{s}_t = \mathbf{A}_t\boldsymbol{\sigma} + \mathbf{n}_t,
\end{equation}
where $\mathbf{A}_t = [\mathbf{a}_{t,1}, \mathbf{a}_{t,2}, \ldots, \mathbf{a}_{t,R}]^{\rm{T}} \in\mathbb{C}^{R\times N}$ is the sensing matrix, and
\begin{equation}
\mathbf{a}_{t,r} = g_t {\rm{diag}}(\mathbf{h}_t^{\rm{e,i}})\mathbf{H}_t^{\rm{i,s}} {\rm{diag}}(\boldsymbol{\omega}_r)\mathbf{h}_t^{\rm{s,p}}.
\end{equation}
Here, $\mathbf{h}_t^{\rm{e,i}}\in\mathbb{C}^{N\times 1}$ and $\mathbf{H}_t^{\rm{i,s}}\in\mathbb{C}^{N\times M}$ have similar forms to $\mathbf{h}_t^{\rm{s,p}}$ in \eqref{eq-h-ris-ap}, indicating the channels from the UE to the ROI and from the ROI to the RIS, respectively. Aggregating the measurements for the $T$ subcarriers, we get
\begin{equation}\label{eq-cs-model-all}
\mathbf{s} = \mathbf{A}\boldsymbol{\sigma} + \mathbf{n},
\end{equation}
where the components are defined as
\begin{equation}
 \mathbf{s} = \left[\begin{matrix} \mathbf{s}_1 \\ \vdots \\ \mathbf{s}_T \end{matrix}\right], ~
 \mathbf{A} = \left[\begin{matrix} \mathbf{A}_1 \\ \vdots \\ \mathbf{A}_T \end{matrix}\right], ~
 \mathbf{n} = \left[\begin{matrix} \mathbf{n}_1 \\ \vdots \\ \mathbf{n}_T \end{matrix}\right].
\end{equation}
The CS-based imaging problem can be described as
\begin{equation}\label{eq-cs-problem-model}
\widehat{\boldsymbol{\sigma}} = \mathop{\arg\min}_{\boldsymbol{\sigma}}  \left\{\|\mathbf{s}-\mathbf{A} {\boldsymbol{\sigma}}\|_2^2 + \beta\|\boldsymbol{\sigma}\|_{1}\right\},
\end{equation}
with $\widehat{\boldsymbol{\sigma}}$ being the estimated image and $\beta$ as the regularization factor. \eqref{eq-cs-problem-model} is solvable using CS-based methods which offer low measurement demands and can enhance sidelobe reduction and noise suppression \cite{zhang2012sparse,fang2013fast,ccetin2014sparsity}.

\begin{algorithm}[t]
\caption{Iterative soft thresholding algorithm (ISTA).}
\label{ag2}
\begin{algorithmic}[1]
        \STATE $\mathbf{input:}$ Sensing matrix $\mathbf{A}$ and extracted measurements $\mathbf{s}$.

        \STATE Initial $\boldsymbol{\sigma}^{(0)}$, $\beta$, $\mu$, and max iteration number $I_{\rm{max}}$.

        \STATE  $\mathbf{for} \  i= 1 \ {\rm{to}} \ I_{\rm{max}} \ \mathbf{do}$

        \STATE \hspace{0.5cm} Derive the intermediate estimate $\mathbf{g}^{(i)}$ based on \eqref{eq-ista-primary}.

        \STATE \hspace{0.5cm} Derive ${\boldsymbol{\sigma}}^{(i)}$ by thresholding $\mathbf{g}^{(i)}$ according to \eqref{eq-ista-thres}.

        \STATE $\mathbf{end\ for}$

        \STATE $\mathbf{output:}$ the estimated ROI image ${\boldsymbol{\sigma}}^{(I_{\max})}$.

\end{algorithmic}
\end{algorithm}

The iterative soft thresholding algorithm (ISTA) \cite{daubechies2004iterative} offers an illustrative comparison point for CS-based imaging techniques when compared with the FT-based imaging approach as mentioned in Algorithm \ref{ag1}. The ISTA progressively refines the estimate of $\boldsymbol{\sigma}$ in every iteration based on
\begin{equation}\label{eq-ista-primary}
\mathbf{g}^{(i)} = \boldsymbol{\sigma}^{(i-1)}+\mu \mathbf{A}^{\rm{H}}\left(\mathbf{s}-\mathbf{A} \boldsymbol{\sigma}^{(i-1)}\right),
\end{equation}
where $\mathbf{g}^{(i)}$ and $\boldsymbol{\sigma}^{(i)}$ symbolize the intermediate and final estimate of $\boldsymbol{\sigma}$ in the $i$-th iteration.
$\mu$ is a convergence-related normalized parameter which adheres to $\mu^{-1} \in (0, \|\mathbf{A}\|_2^2)$.
To ensure the sparsity of the solution, a thresholding operator ${\rm{E}}_{\chi}(\cdot)$ is applied to $\mathbf{g}^{(i)}$. The outcome of this operation in the $i$-th iteration is given as
\begin{equation}\label{eq-ista-thres}
\boldsymbol{\sigma}^{(i)}={\rm{E}}_{\chi}{\left(\mathbf{g}^{(i)}\right)}.
\end{equation}
The thresholding operation is characterized by
\begin{equation}\label{threshold}
{\rm{E}}_{\chi} {\left(g_n^{(i)}\right)} = \left\{\begin{array}{ll}
 {\rm{sgn}} {\left(g_n^{(i)}\right)} {\left(|g_n^{(i)}|-\chi\right)}, &  \text { if }|g_n^{(i)}|  \geq \chi, \\
 0, &  \text { otherwise}.
\end{array}\right.
\end{equation}
where $g_n^{(i)}$ is the $n$-th element of $\mathbf{g}^{(i)}$.
Here, the selection of $\chi=\beta\mu$ typically depends on the sparsity of $\boldsymbol{\sigma}$ and is determined by the magnitude of the $(\alpha+1)$-th largest component of $\mathbf{g}$ with $\alpha$ being the sparsity of $\boldsymbol{\sigma}$.
The ISTA's specifics are highlighted in Algorithm \ref{ag2}.

When $R=M$ and $N=MN_{\rm{x}}$, the computational complexities of Algorithms \ref{ag1} and \ref{ag2} can be given as $O(TM{\rm{log}}M)$ and $O(I_{\rm{max}}TM^2N_{\rm{x}})$, respectively.
Therefore, the computational burden of Algorithm \ref{ag2} is significantly higher than that of Algorithm \ref{ag1}.
Notably, CS-based techniques require the calculation of the sensing matrix, which involves computations of $\mathbf{h}_t^{\rm{e,i}}$, $\mathbf{H}_t^{\rm{i,s}}$, and $\mathbf{h}_t^{\rm{s,p}}$. These not only take significant time but also demand substantial computational resources.
While the FT-based Algorithm \ref{ag1} stands out with its efficiency and minimal memory requirement, a notable drawback is the unavoidable sidelobes in its outcome. However, these sidelobes can be substantially curtailed in CS-based methodologies. Table \ref{tab-ft-cs-comp} summarizes the performance comparison between FT- and CS-based imaging techniques. The choice between them, in practice, should be balanced among various factors, including available computational resources, memory constraints, and measurement numbers.

\begin{table}[t]
  \renewcommand{\arraystretch}{1.5}
  \centering
  \fontsize{8}{8}\selectfont
  \captionsetup{font=small}
  \caption{Comparison between FT- and CS-based imaging methods.}\label{tab-ft-cs-comp}
  \begin{threeparttable}
    \begin{tabular}{p{2.1cm}<{\centering}|p{2.7cm}<{\centering}|p{2.7cm}<{\centering}}
      \specialrule{1pt}{0pt}{-1pt}\xrowht{10pt}
      Comparison & FT-based imaging & CS-based imaging \\
      \hline
      \multirow{2}{*}{System parameters} & \multicolumn{2}{c}{RIS phase shift configuration matrix} \\[-4pt]
       & \multicolumn{2}{c}{and the locations of the UE, AP, RIS, and ROI} \\
      \hline
      Prior information & / & Sparsity \\
      \hline
      Measurement & Satisfying the & Smaller than the \\[-4pt]
      number & Nyquist criterion & Nyquist criterion \\
      \hline
      Preparation & \multirow{2}{*}{/} & Sensing matrix \\[-4pt]
      before imaging & & generation \\
      \hline
      Computational & \multirow{2}{*}{$O(TM(\log_2{M} +N_{\rm{x}}))$} & \multirow{2}{*}{$O(I_{\rm{max}}TRN)$} \\[-4pt]
      complexity & & \\
      \hline
      Memory cost & Low & High \\
      \hline
      Sidelobes & Unavoidable & Reduced \\
      \hline
      Background noise & Reserved & Suppressed \\
      \hline
      Parallel  & \multirow{2}{*}{Supportable} & \multirow{2}{*}{Unsupportable} \\[-4pt]
      computation & & \\
      \specialrule{1pt}{0pt}{-1pt}
    \end{tabular}
  \end{threeparttable}
\end{table}

\subsection{Integrated FT and CS Imaging Algorithm}
\label{sec-integrated-ft-cs-imaging}

As outlined in Sec. \ref{sec-ft-cs-compare}, both FT and CS imaging algorithms offer distinct benefits in terms of computational complexity, memory expense, measurement count, and image quality. However, when resource constraints are present in communication systems, neither algorithm may be ideal for practical implementation. This subsection proposes an integrated approach, leveraging the strengths of both FT- and CS-based methods, to achieve superior imaging quality with reduced time, memory, and measurement overheads.

According to Algorithm \ref{ag2}, significant computation and memory overheads arise from the creation of the large-dimensional sensing matrix and its multiplication with vectors. The core integration strategy for FT and CS algorithms involves substituting the sensing matrix with FT-based imaging operators, which can be effectively realized using a combination of 1D FFT, IFFT, and Hadamard operations \cite{fang2013fast,bi2019wavenumber}. This bypasses the need for generating the sensing matrix, and matrix-vector products can be efficiently executed using FFT and IFFT processes. Drawing inspiration from \cite{fang2013fast,bi2019wavenumber}, we introduce a forward operator $\boldsymbol{\Re}$ and a backward operator $\boldsymbol{\Im}$ to supplant $\mathbf{A}$ and $\mathbf{A}^{\rm{H}}$ in Algorithm \ref{ag2}, respectively.

Let $\widetilde{\boldsymbol{\sigma}}_t$ represent the 3D subimage generated by Algorithm \ref{ag1} based on the noise-free measurements $\mathbf{s}_t^\circ$ at the $t$-th subcarrier, where $\mathbf{s}_{t}^\circ = g_t \boldsymbol{\Omega}_R (\mathbf{b}_t \odot \mathbf{h}_t^{\rm{s,p}})$ according to \eqref{eq-min-new-vec}. When $R = M$, the imaging procedure transforms $\mathbf{s}_t^\circ$ to the subimage $\widetilde{\boldsymbol{\sigma}}_t$ using the backward sub-operator $\boldsymbol{\Im}_t$, expressed as
\begin{equation}
\widetilde{\boldsymbol{\sigma}}_t = \boldsymbol{\Im}_t(\mathbf{s}_t^\circ).
\end{equation}
Assuming $\boldsymbol{\Omega}_R$ as the DFT matrix, $\boldsymbol{\Im}_t$ is derived from \eqref{eq-recover-bt-fft}, \eqref{eq-b'2}, \eqref{eq-sigma-t-estimate}, and \eqref{eq-sigma-t-estimate-new} in discretized forms, given as
\begin{equation}\label{eq-backward-operator}
\begin{aligned}
& \quad \boldsymbol{\Im}_t(\mathbf{s}_t^\circ) \\
& = {\cal F}^{-1}_{k_{y}, k_{z}}  \Bigg\{
  {\cal F}_{y, z} \Big\{
  \boldsymbol{\aleph}(\underbrace{  {\cal F}^{-1} \{\mathbf{s}_{t}^\circ\}  \oslash  (g_t\mathbf{h}_{t}^{\rm{s,p}})}_{\text{ECR recovery}})
  \Big\}  \odot \mathbf{P}_{t,1}  \Bigg\}  \odot  \mathbf{P}_{t,2},
\end{aligned}
\end{equation}
where $\boldsymbol{\aleph}(\cdot)$ reshapes and repeats the data to match the matrix dimensions.
The element of $\mathbf{P}_{t,1}\in\mathbb{C}^{N_{\rm{x}}\times M_{\rm{y}}\times M_{\rm{z}}}$ at $(n_{\rm{x}}, k_y, k_z)$ is presented as
\begin{equation}
\mathbf{P}_{t,1}(n_{\rm{x}}, k_y, k_z) = -jk_xe^{jk_xD_0}e^{jk_xx_{n_{\rm{x}}}},
\end{equation}
with $x_{n_{\rm{x}}}$ denoting the location of the $n_{\rm{x}}$-th sampling point along the x-axis in \eqref{eq-sigma-t-estimate}.
In contrast, the $(n_{\rm{x}}, n_{\rm{y}}, n_{\rm{z}})$-th element of $\mathbf{P}_{t,2}\in\mathbb{C}^{N_{\rm{x}}\times N_{\rm{y}}\times N_{\rm{z}}}$ is given as
\begin{equation}
\mathbf{P}_{t,2}(n_{\rm{x}}, n_{\rm{y}}, n_{\rm{z}}) = \frac{1}{2d_1}e^{-jk_td_1},
\end{equation}
where $d_1$ represents the distance from the UE to the voxel in the ROI.
The dimensions of both $\mathbf{P}_{t,1}$ and $\mathbf{P}_{t,2}$ are influenced by the RIS element number, the last IDFT size of \eqref{eq-backward-operator}, and x-axis sampling numbers.
Given $\mathbf{s}^\circ = [\mathbf{s}_1^\circ; \mathbf{s}_2^\circ; \ldots; \mathbf{s}_T^\circ]$, the final ROI image is formulated as
\begin{equation}\label{eq-backward-operator-final}
\widetilde{\boldsymbol{\sigma}} = \boldsymbol{\Im}(\mathbf{s}^\circ) = \sum_{t=1}^T\widetilde{\boldsymbol{\sigma}}_t = \sum_{t=1}^T\boldsymbol{\Im}_t(\mathbf{s}_t^\circ).
\end{equation}

Assuming $\boldsymbol{\Im}$ sufficiently reconstructs the ROI image with minimal sidelobes, such that $\widetilde{\boldsymbol{\sigma}}$ closely represents $\boldsymbol{\sigma}$, we get $\mathbf{A}\boldsymbol{\Im}(\mathbf{s}^\circ) \approx \mathbf{A}{\boldsymbol{\sigma}} = \mathbf{s}^\circ$ according to \eqref{eq-cs-model-all}. This leads to the approximation
\begin{equation}
\mathbf{s}^\circ \approx \boldsymbol{\Im}^{-1}(\widetilde{\boldsymbol{\sigma}}) = \boldsymbol{\Re}(\widetilde{\boldsymbol{\sigma}}),
\end{equation}
where $\boldsymbol{\Re}$ denotes the forward operator converting $\widetilde{\boldsymbol{\sigma}}$ to $\mathbf{s}^\circ$.
However, deriving $\boldsymbol{\Re}$ directly is challenging due to the irreversible accumulation operation in \eqref{eq-backward-operator-final}. But, the reversible backward sub-operator in \eqref{eq-backward-operator} allows us to deduce the forward sub-operator $\boldsymbol{\Re}_t$ by inverting each step in \eqref{eq-backward-operator}.
Hence, we approximate $\boldsymbol{\Re}$ with $\{\boldsymbol{\Re}_t\}_{t=1}^T$, formulated as
\begin{equation}
\boldsymbol{\Re}(\widetilde{\boldsymbol{\sigma}}) \approx \left[ \begin{matrix} \boldsymbol{\Re}_1(\widetilde{\boldsymbol{\sigma}}) \\ \vdots \\ \boldsymbol{\Re}_T(\widetilde{\boldsymbol{\sigma}}) \end{matrix} \right].
\end{equation}
The linear signal model \eqref{eq-cs-model-all} can then be expressed as
\begin{equation}
\mathbf{s} = \mathbf{A}\boldsymbol{\sigma}+\mathbf{n} \approx \boldsymbol{\Re}(\widetilde{\boldsymbol{\sigma}})+\mathbf{n}.
\end{equation}
\hspace{-0.5em}Thus, we propose to reconstruct $\widetilde{\boldsymbol{\sigma}}$ using the approximated forward operator $\boldsymbol{\Re}$. Based on \cite{fang2013fast}, the relationship between $\widetilde{\boldsymbol{\sigma}}$ and ${\boldsymbol{\sigma}}$ can be formulated as
\begin{equation}\label{epsilon}
\widetilde{\boldsymbol{\sigma}} = \boldsymbol{\sigma} \odot \mathbf{p} + \boldsymbol{\epsilon},
\end{equation}
where $\mathbf{p}$ represents phase errors and $\boldsymbol{\epsilon}$ covers sidelobes and artifacts.
If the FT-based imaging operator $\boldsymbol{\Im}$ is finely tuned for optimal focus and the sidelobes in $\widetilde{\boldsymbol{\sigma}}$ are minimal, the error term 
$\boldsymbol{\epsilon}$ becomes negligible. This suggests that $\widetilde{\boldsymbol{\sigma}}$ and ${\boldsymbol{\sigma}}$ preserve comparable levels of sparsity. Consequently, by applying the operators $\boldsymbol{\Re}$ and $\boldsymbol{\Im}$, we are able to reconstruct a closely approximate version of ${\boldsymbol{\sigma}}$ with $\widetilde{\boldsymbol{\sigma}}$.

In Algorithm \ref{ag2}, $\mathbf{A}$ maps the estimate $\boldsymbol{\sigma}^{(i)}$ to its corresponding channel measurements, mirroring the role of the operator $\boldsymbol{\Re}$. Conversely, $\mathbf{A}^{\rm{H}}$ reconstructs the image from the measurements, akin to the function of the operator $\boldsymbol{\Im}$.
Consequently, to integrate the FT- and CS-based imaging methods, we suggest substituting matrix-vector multiplications $\mathbf{A}\boldsymbol{\sigma}^{(i)}$ and $\mathbf{A}^{\rm{H}}\mathbf{{r}}^{(i)}$ in Algorithm \ref{ag2} with $\boldsymbol{\Re}(\boldsymbol{\sigma}^{(i)})$ and $\boldsymbol{\Im}(\mathbf{r}^{(i)})$, respectively.
Here, $\mathbf{r}^{(i)}=\mathbf{s}-\mathbf{A} \boldsymbol{\sigma}^{(i)}$ denotes the residual measurements in the $i$-th iteration. However, determining $\|\mathbf{A}\|^2_2$ becomes challenging with approximated operators. Hence, we employ the adaptive step selection strategy \cite{blumensath2010normalized} to ascertain the normalized parameter $\mu$ during the $i$-th iteration, given as
\begin{equation}\label{eq-normalized-parameter}
\mu^{(i)} =  \left\|\Delta \boldsymbol{\sigma}_{\alpha}^{(i)} \right\|_{2}^{2}  / \left\|\boldsymbol{\Re}(\Delta \boldsymbol{\sigma}_{\alpha}^{(i)}) \right\|_{2}^{2},
\end{equation}
where $\Delta \boldsymbol{\sigma}_{\alpha}^{(i)}$ is equivalent to $\Delta \boldsymbol{\sigma}^{(i)}=\boldsymbol{\Im}(\mathbf{r}^{(i-1)})$ on the support of $\boldsymbol{\sigma}^{(i-1)}$ and 0 elsewhere. The integrated imaging algorithm of FT and CS is outlined in Algorithm \ref{ag3}. The computational complexity per iteration drops from $O(TM^2N_{\rm{x}})$ to $O(TM\log_2M)$ for $R = M$ and $N = MN_{\rm{x}}$.

For cases with limited measurements, one can employ Algorithm \ref{ag3} in conjunction with techniques introduced in Sec. \ref{sec-preliminary}. Specifically, when using the pseudo-inverse matrix approach, the sensing matrix can be approximated as $\boldsymbol{\Theta\Re}$. Here, the diagonal elements of $\boldsymbol{\Theta}\in\mathbb{R}^{R\times M}$ are set to one, while other elements are zero. This samples out the sensing matrix associated with the initial $R$ measurements. On the other hand, when using the block-controlled RIS method, the data dimension for ECR recovery in $\boldsymbol{\Re}$ is reduced to $M_Q$. Zero-padding can then be applied during IDFT operations to reduce the voxel size.

\begin{algorithm}[t]
\caption{Integrated imaging algorithm of FT and CS.}
\label{ag3}
\begin{algorithmic}[1]
        \STATE $\mathbf{input:}$ Extracted measurements $\mathbf{s}$.

        \STATE Initial $\boldsymbol{\sigma}^{(0)}$, $\beta$, $\mu$, and max iteration $I_{\rm{max}}$.

        \STATE  $\mathbf{for} \  i= 1 \ {\rm{to}} \ I_{\rm{max}} \ \mathbf{do}$

        \STATE \hspace{0.5cm} Calculate the residual signal by $\mathbf{r}^{(i-\!1)}\!=\!\mathbf{s}-\boldsymbol{\Re}\!\left(\!\boldsymbol{\sigma}^{(i-\!1)}\!\right)$.

        \STATE \hspace{0.5cm} Project the residual signal to the image domain by
        \\ \hspace{1.0cm} $\Delta\boldsymbol{\sigma}^{(i)} = \boldsymbol{\Im}(\mathbf{r}^{(i-1)})$.

        \STATE \hspace{0.5cm} Calculate the adaptive normalized parameter $\mu^{(i)}$
        \\ \hspace{1.0cm} according to \eqref{eq-normalized-parameter}.

        \STATE \hspace{0.5cm} Threshold by $\boldsymbol{\sigma}^{(i)}={\rm{E}}_{\chi}\left( \boldsymbol{\sigma}^{(i-1)} + \mu^{(i)}\Delta \boldsymbol{\sigma}^{(i)} \right)$.

        \STATE $\mathbf{end\ for}$

        \STATE $\mathbf{output:}$ the estimated ROI image ${\boldsymbol{\sigma}}^{(I_{\max})}$.

\end{algorithmic}
\end{algorithm}

\begin{remark}
Algorithm \ref{ag3} effectively merges the strengths of both FT- and CS-based methods. Initially, it builds upon the ISTA, maintaining its benefits of sidelobe reduction, noise suppression, and enabling superior imaging even when $R<M$. Additionally, the tasks of sensing matrix generation and matrix-vector multiplication in the traditional ISTA are supplanted by FT-based operators. This results in significant savings in computational complexity and memory cost compared to Algorithm \ref{ag2}. Thus, Algorithm \ref{ag3} delivers high-quality imaging with fewer time, memory, and measurement demands.
\end{remark}

\begin{remark}
Algorithm \ref{ag3} distinctly diverges from previous studies \cite{fang2013fast,bi2019wavenumber}. At its core, it is crafted for radio imaging within RIS-aided communication systems, leveraging OFDM pilots for measurements. In stark contrast, \cite{fang2013fast,bi2019wavenumber} focus on imaging in specific radar setups using FMCW signals. The constructed operators, $\boldsymbol{\Re}$ and $\boldsymbol{\Im}$, also differ notably. First, they are formed using an interpolation-free wavenumber domain algorithm tailored for near-field bistatic systems, while the aforementioned studies employ far-field monostatic methods requiring interpolation. Secondly, our operators do not solely rely on the traditional SAA; they also incorporate the ECR recovery algorithm as seen in \eqref{eq-recover-bt-fft}. Finally, given the irreversibility of the subimage accumulation operation in $\boldsymbol{\Im}$, we approximate $\boldsymbol{\Re}$ using a series of sub-operators $\{\boldsymbol{\Re}_t\}_{t=1}^T$, each catering to an individual frequency.
\end{remark}

\section{Numerical Results}

In this section, we assess the performance of our proposed algorithms through simulations.
Unless otherwise specified, all coordinates, distances, and dimensions are normalized by the center frequency's wavelength $\lambda_0$.
The RIS is comprised of $100\times 100$ elements, each size of $\frac{1}{2}\times \frac{1}{2}$. It is centered at $[-50, 0, 0]^{\rm{T}}$, with phase shifts set according to the DFT matrix. The AP and UE are situated at $[20, 20, 20]^{\rm{T}}$ and $[-40, -40, 0]^{\rm{T}}$, respectively, while the ROI's center is at the origin.
The bandwidth and subcarrier spacing are expressed as ratios to the center frequency $f_0$: $\eta_B = B / f_0 = 1 / 15$ and $\eta_{\Delta f} = \Delta f / f_0 = 1 / 300$, respectively.
Consequently, $T=21$ subcarriers are employed for imaging, ensuring that the subcarrier spacing $\Delta f$ meets the Nyquist sampling criterion.
When the bandwidth is constant, augmenting the number of subcarriers can diminish noise in the imaging outcomes but may not enhance the range resolution.

\subsection{3D Imaging Results of Algorithm \ref{ag1} for a Point Target}
\label{sec-result-1}

\begin{figure}
\centering
\captionsetup{font=footnotesize}
\begin{subfigure}[b]{0.52\linewidth}
\centering
\includegraphics[width=\linewidth]{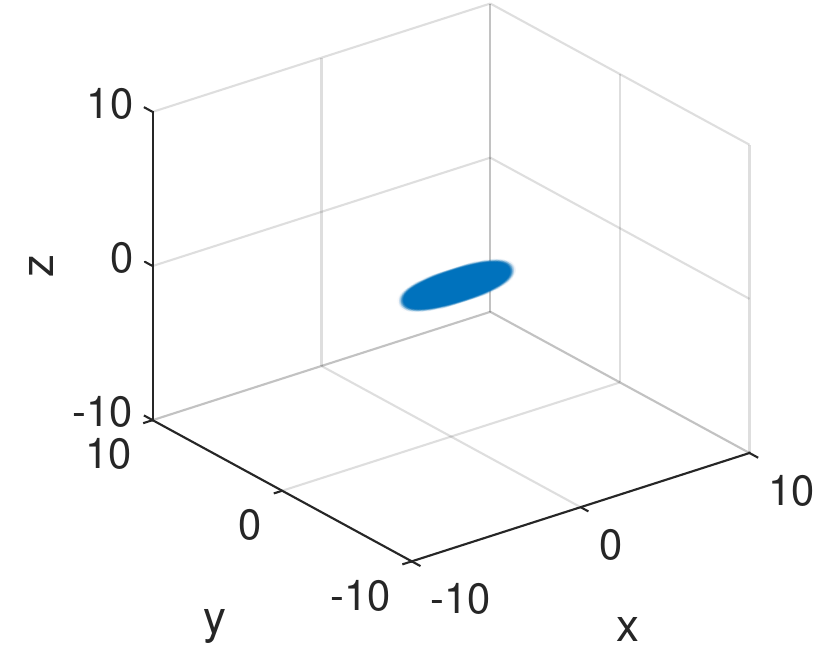}
\caption{}
\label{fig-result-1-a}
\end{subfigure}
\begin{subfigure}[b]{0.48\linewidth}
\centering
\includegraphics[width=\linewidth]{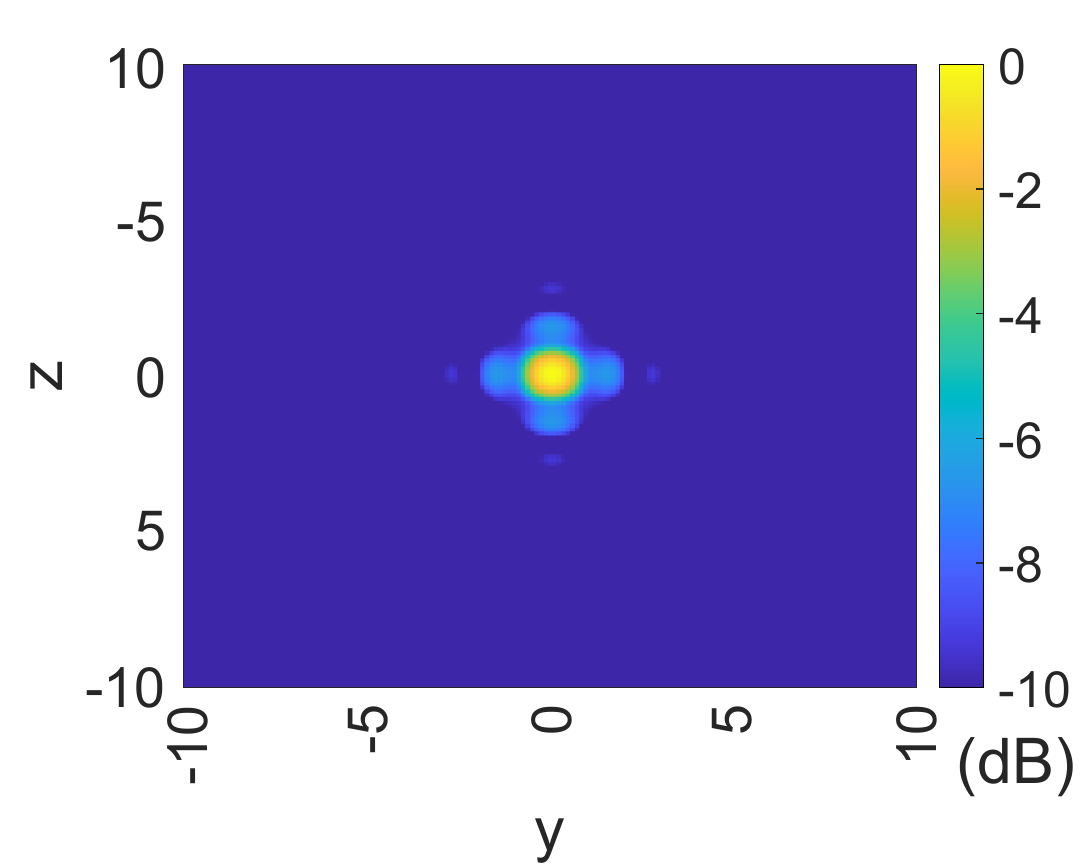}
\caption{}
\label{fig-result-1-b}
\end{subfigure}
\begin{subfigure}[b]{0.48\linewidth}
\centering
\includegraphics[width=\linewidth]{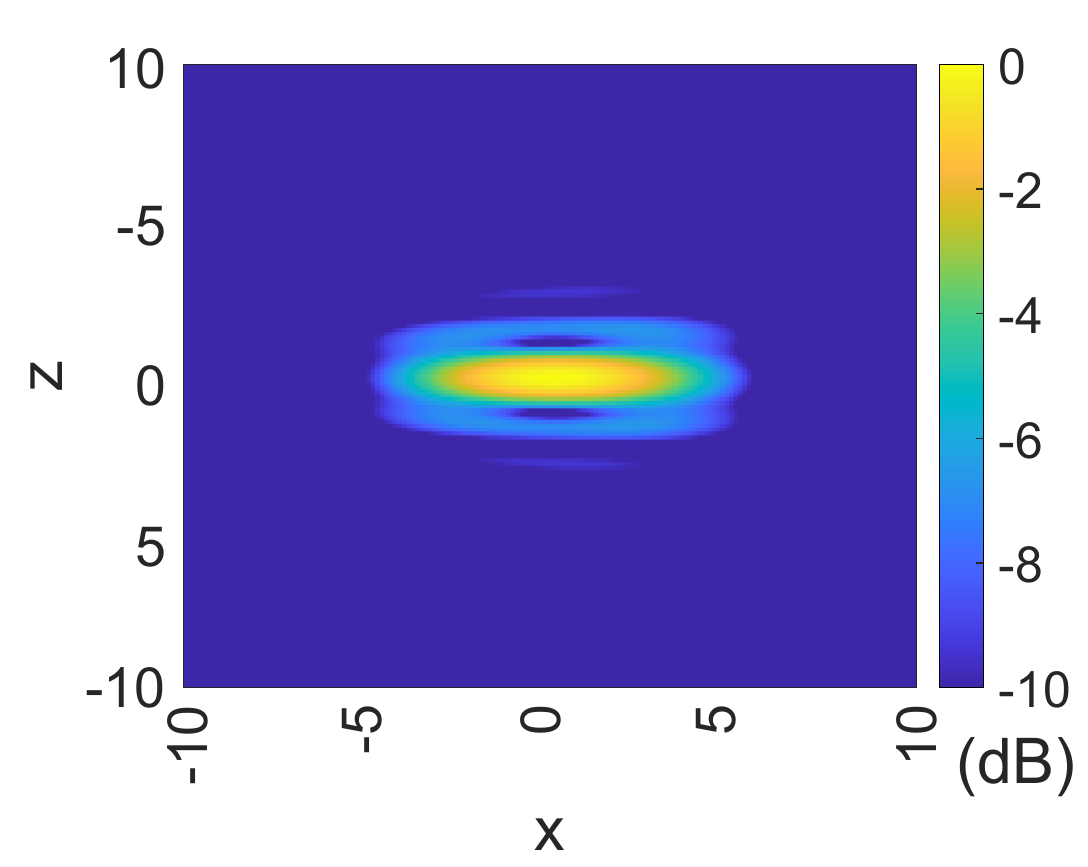}
\caption{}
\label{fig-result-1-c}
\end{subfigure}

\begin{subfigure}[b]{0.42\linewidth}
\centering
\includegraphics[width=\linewidth]{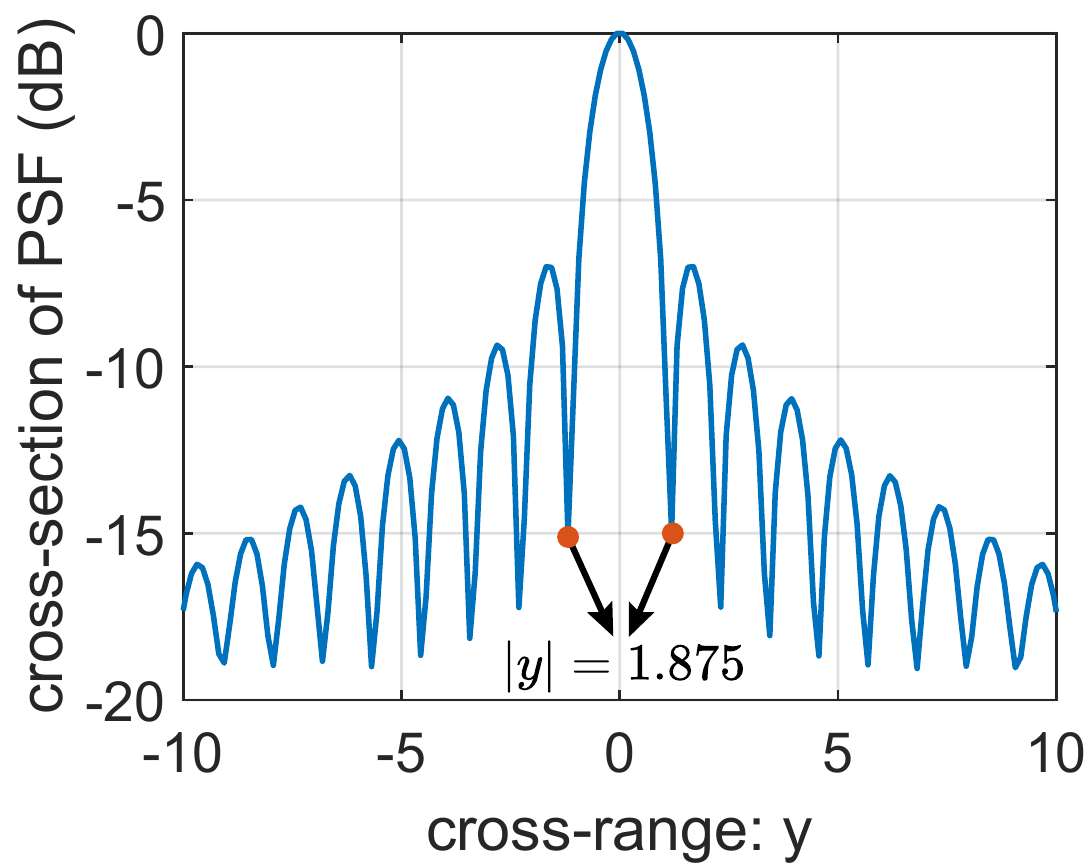}
\caption{}
\label{fig-result-1-d}
\end{subfigure}
\quad\quad
\begin{subfigure}[b]{0.42\linewidth}
\centering
\includegraphics[width=\linewidth]{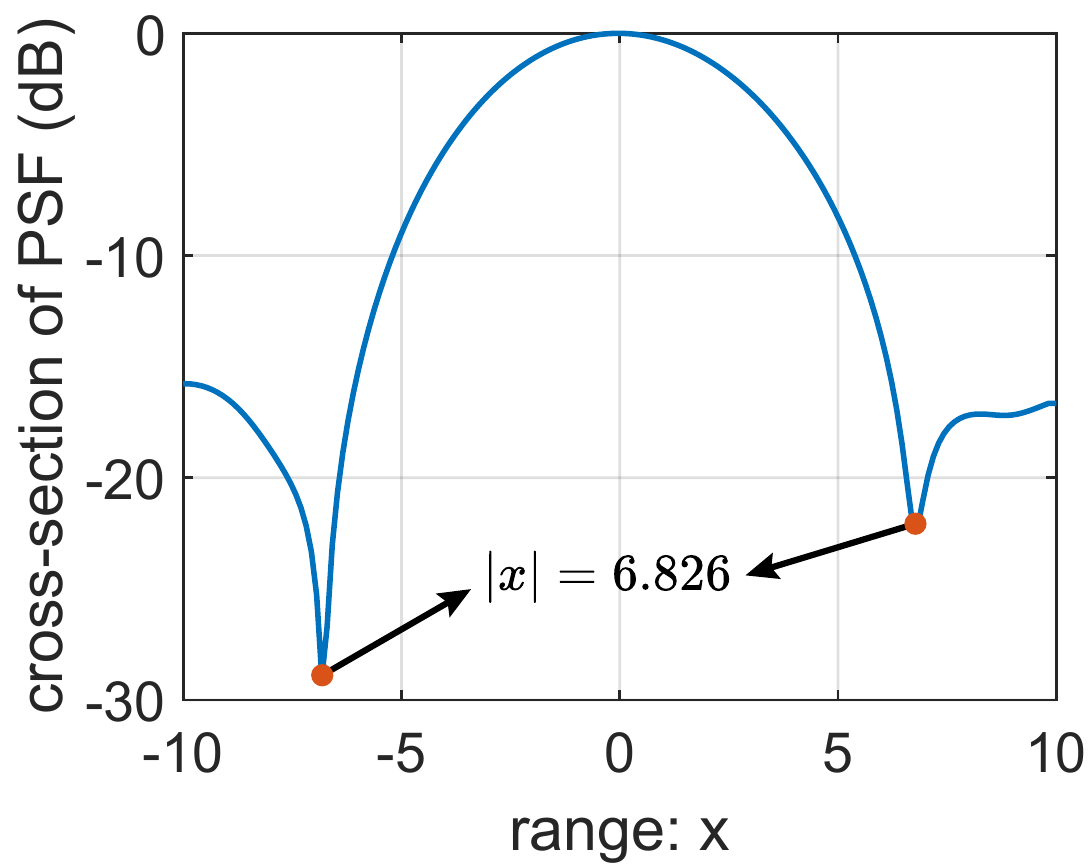}
\caption{}
\label{fig-result-1-e}
\end{subfigure}

\caption{Imaging results of Algorithm \ref{ag1} for a single point target: (a) 3D imaging result; (b) 2D projection into the y-z plane; (c) 2D projection into the x-z plane; (d) 1D PSF along the y-axis; (e) 1D PSF along the x-axis.}
\label{fig-result-1}
\end{figure}

This subsection evaluates the effectiveness of the proposed FT-based two-step wavenumber domain imaging Algorithm \ref{ag1}. The imaging results of a single point target are illustrated in Fig. \ref{fig-result-1}.
The region of the ROI is given as $\mathbb{D} = \{[x, y, z]^{\rm{T}} : {-10 \le x \le 10}, {-10 \le y \le 10}, {-10 \le z \le 10} \}$, where a single point target is located at the origin with a unit scattering coefficient.
The energy of the UE-ROI-RIS-AP path is normalized, and the additive noise variance in the measurements is set to 0.01.
Fig. \ref{fig-result-1-a} depicts the 3D imaging result of the point target, while Figs. \ref{fig-result-1-b} and \ref{fig-result-1-c} are the 2D projections into the y-z and x-z planes, respectively.
These three figures demonstrate that the proposed Algorithm \ref{ag1} can effectively formulate the ROI image and accurately detect the single point target at the origin. However, the imaging results display sidelobes around the target, which are further shown in Figs. \ref{fig-result-1-d} and \ref{fig-result-1-e} by plotting the 1D cross-sections of the PSF along the y- and x-axis, respectively. As derived in \eqref{eq-kspace-ifft}, the PSF cross-sections possess the shape of the ${\rm{Sa}}(\cdot)$ function. Moreover, the numerical resolution in the considered scenario can be obtained from Figs. \ref{fig-result-1-d} and \ref{fig-result-1-e}, which is $1.875$ in the cross-range direction and $6.826$ in the range direction. The cross-range resolution is better than that in the range direction due to the large RIS-subtended angle and limited communication bandwidth, as revealed by \eqref{eq-kspace-limit-x} and \eqref{eq-kspace-limit-y}.

\begin{figure}
\centering
\captionsetup{font=footnotesize}
\quad\ \begin{subfigure}[b]{0.84\linewidth}
\centering
\includegraphics[width=\linewidth]{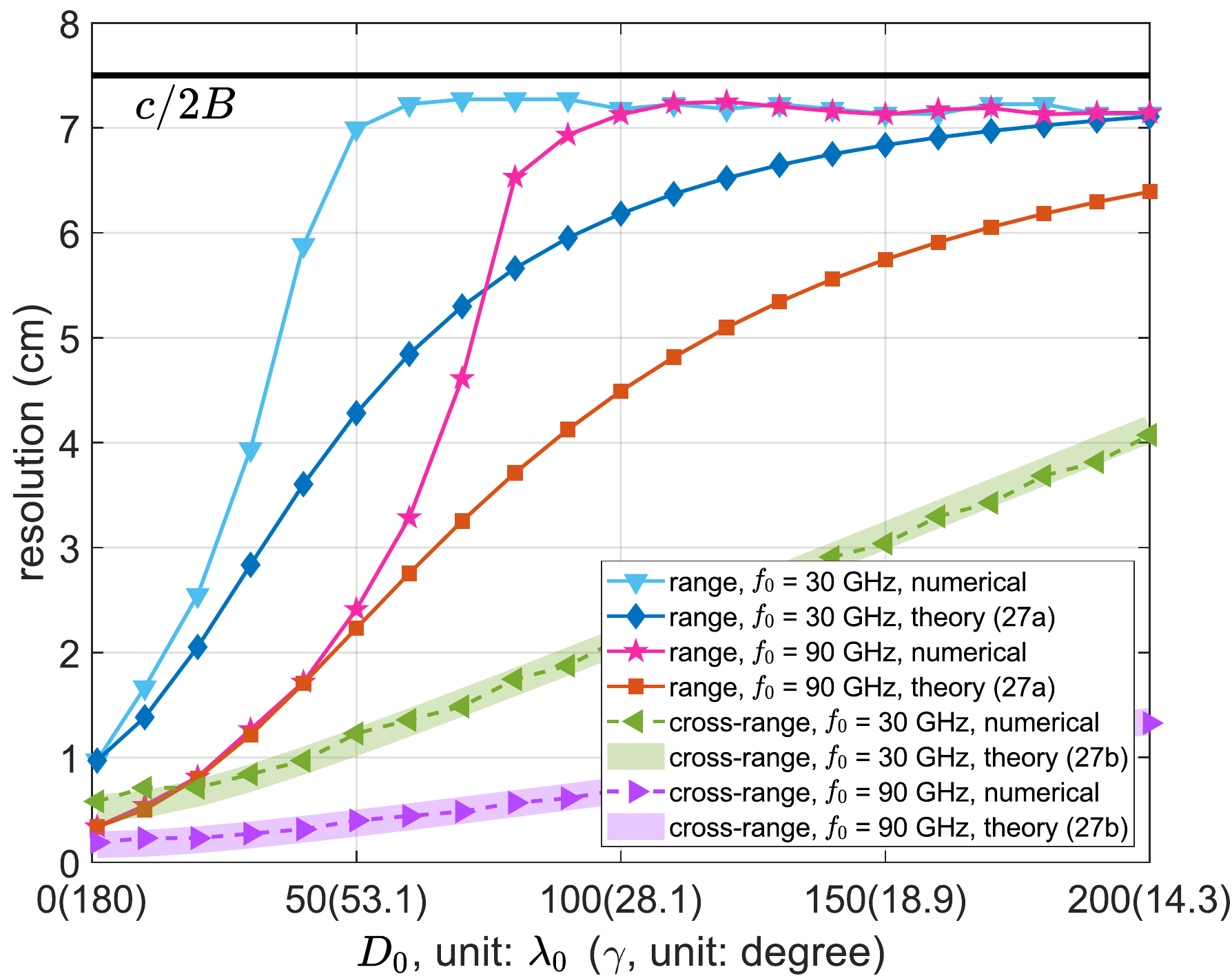}
\caption{Resolutions versus the distance $D_0$ (the angle $\gamma$).}
\label{fig-result-2-a}
\end{subfigure}
\begin{subfigure}[b]{0.8\linewidth}
\centering
\includegraphics[width=\linewidth]{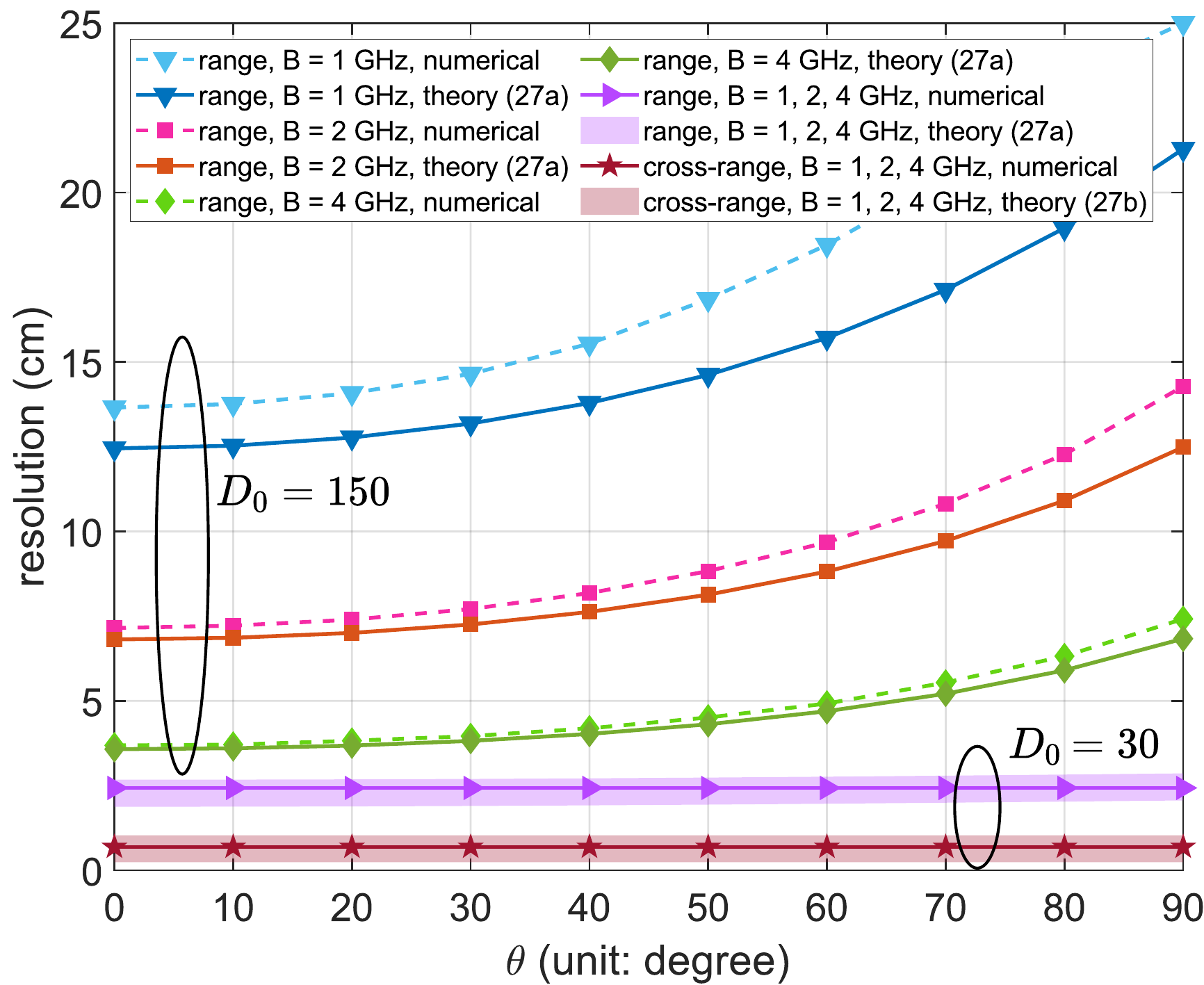}
\caption{Resolutions versus the angle $\theta$.}
\label{fig-result-2-b}
\end{subfigure}

\caption{Comparisons of theoretical and numerical resolution limits.}
\label{fig-result-2}
\end{figure}

\subsection{Comparison of Numerical Resolutions and the DRLs}
\label{sec-result-2}

This subsection evaluates the performance of the proposed Algorithm \ref{ag1} by comparing the numerical resolutions and the proposed DRLs in Sec. \ref{sec-drl}.
The simulation scenario is the same as that in Sec. \ref{sec-result-1}, and the numerical results are derived from the PSF of the imaging results of Algorithm \ref{ag1}, similar to Figs. \ref{fig-result-1-d} and \ref{fig-result-1-e}.
The theoretical results are calculated by \eqref{eq-kspace-limit-x} and \eqref{eq-kspace-limit-y}.
Moreover, the impacts of various system parameters on the resolutions are discussed, using absolute frequency, bandwidth, and resolution values.

First, we compare the numerical and theoretical resolutions with varying distance $D_0$ (equivalently, the RIS-subtended angle $\gamma$) and working frequency $f_0$ in Fig. \ref{fig-result-2-a}.
The bandwidth $B = 2\ \text{GHz}$, and the transmitting direction $\theta = 0^\circ$, resulting in a constant traditional range resolution of $c/2B$.
The simulation results demonstrate that the range resolution value is positively correlated with the distance $D_0$, and the numerical resolutions of Algorithm \ref{ag1} approach the DRLs when $D_0$ is small, verifying the effectiveness of this algorithm.
Moreover, the increase in $f_0$ can degrade the range resolution value, as revealed by \eqref{eq-kspace-limit-x}.
However, the numerical range resolutions approach the traditional DRL of $c/2B$ when $D_0$ is relatively large, since the second term of the denominator in \eqref{eq-kspace-limit-x} approaches zero.
The proposed theoretical range DRL and the traditional value of $c/2B$ act as the lower and upper bounds of the numerical results, respectively.
Additionally, the numerical cross-range resolution of Algorithm \ref{ag1} matches well with the DRL \eqref{eq-kspace-limit-y} and grows with the increase in $D_0$ and the decrease in $f_0$.

Second, the numerical and theoretical results are compared with various bandwidth $B$ and transmitting direction $\theta$ in Fig. \ref{fig-result-2-b}, where $f_0 = 30\ \text{GHz}$ and $D_0 = 150$ or $D_0 = 30$.
In contrast to the traditional DRL of constant $c/2B$, the simulation results have shown that the numerical and theoretical range resolution values increase together with $\theta$ growing when $D_0$ is large.
Moreover, the range resolution decreases with the increase in $B$.
Meanwhile, the gap between the numerical and theoretical values also degrades, meaning that Algorithm \ref{ag1} obtains nearly the optimal imaging performance under the Rayleigh criterion.
However, when $D_0$ is small (e.g., $D_0=30$ in Fig. \ref{fig-result-2-b}), the variation of $B$ leads to minimal effects on the range resolution because $\gamma$ is large, and $B$ is much smaller than $f_0$, resulting in the dominance of the second term in the denominator of \eqref{eq-kspace-limit-x}.
Finally, the simulation results have also validated that the cross-range resolution is not relevant to the angle $\theta$ in equivalent SIMO bistatic imaging systems.

\subsection{Imaging Results of Algorithm \ref{ag1} with Different RIS Array Sizes and Discrete RIS Phase Shifts}
\label{sec-result-3}

\begin{figure}
\centering
\captionsetup{font=footnotesize}
\begin{subfigure}[b]{0.485\linewidth}
\centering
\includegraphics[width=\linewidth]{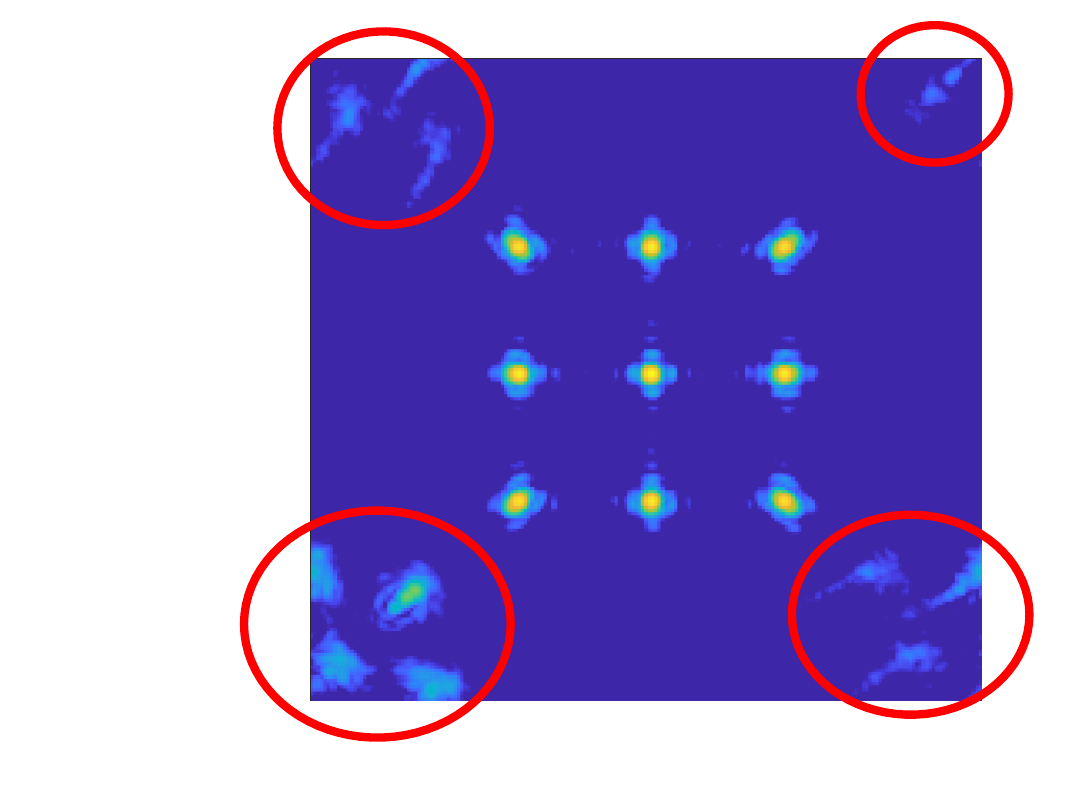}
\vspace{-0.8cm}
\caption{}
\label{fig-result-3-a}
\end{subfigure}
\begin{subfigure}[b]{0.4\linewidth}
\centering
\includegraphics[width=\linewidth]{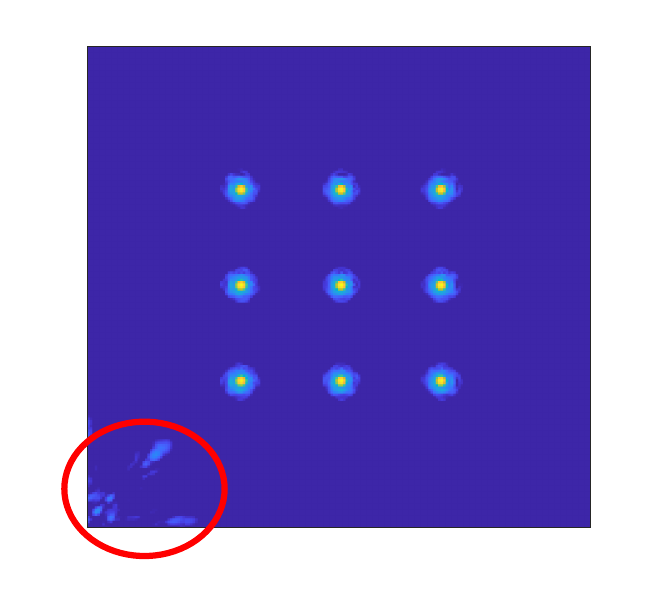}
\vspace{-0.8cm}
\caption{}
\label{fig-result-3-b}
\end{subfigure}

\begin{subfigure}[b]{0.485\linewidth}
\centering
\includegraphics[width=\linewidth]{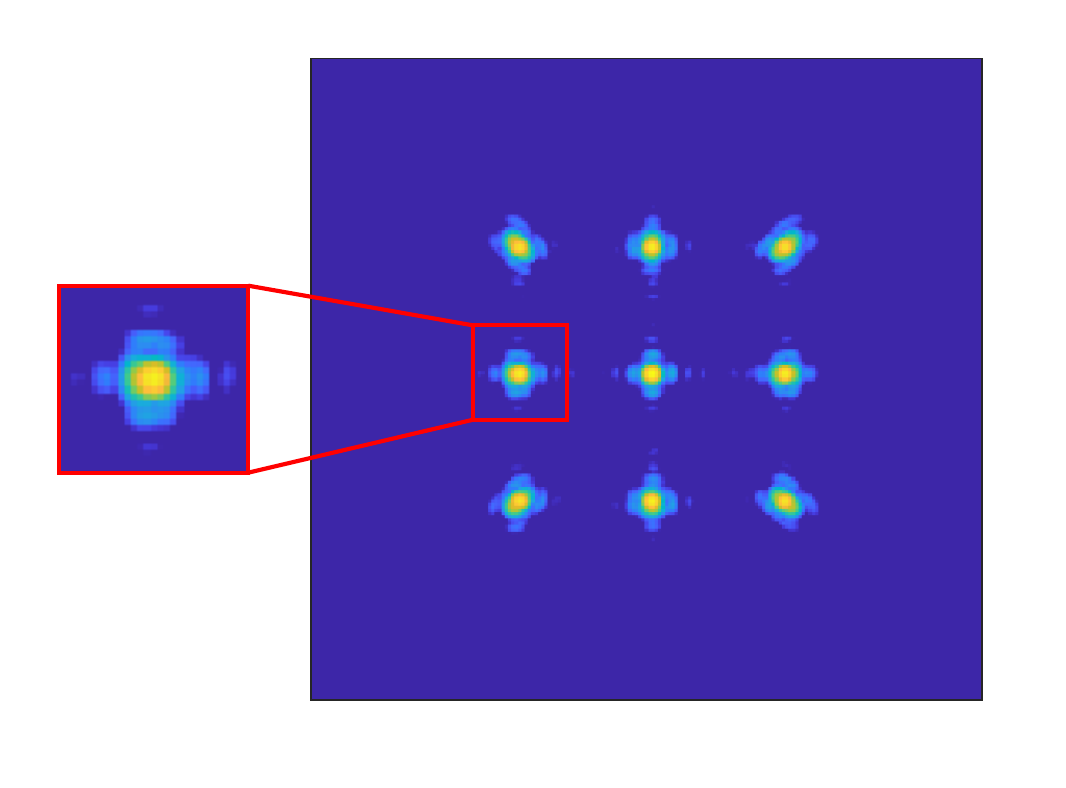}
\vspace{-0.8cm}
\caption{}
\label{fig-result-3-c}
\end{subfigure}
\begin{subfigure}[b]{0.4\linewidth}
\centering
\includegraphics[width=\linewidth]{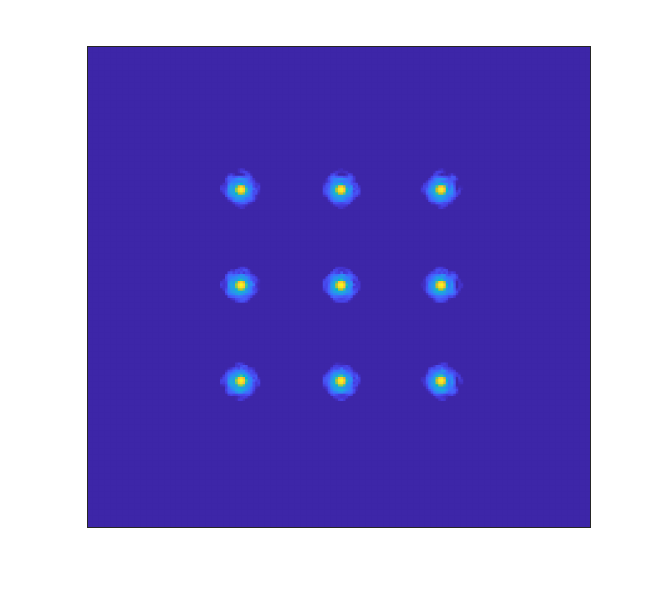}
\vspace{-0.8cm}
\caption{}
\label{fig-result-3-d}
\end{subfigure}

\begin{subfigure}[b]{0.485\linewidth}
\centering
\includegraphics[width=\linewidth]{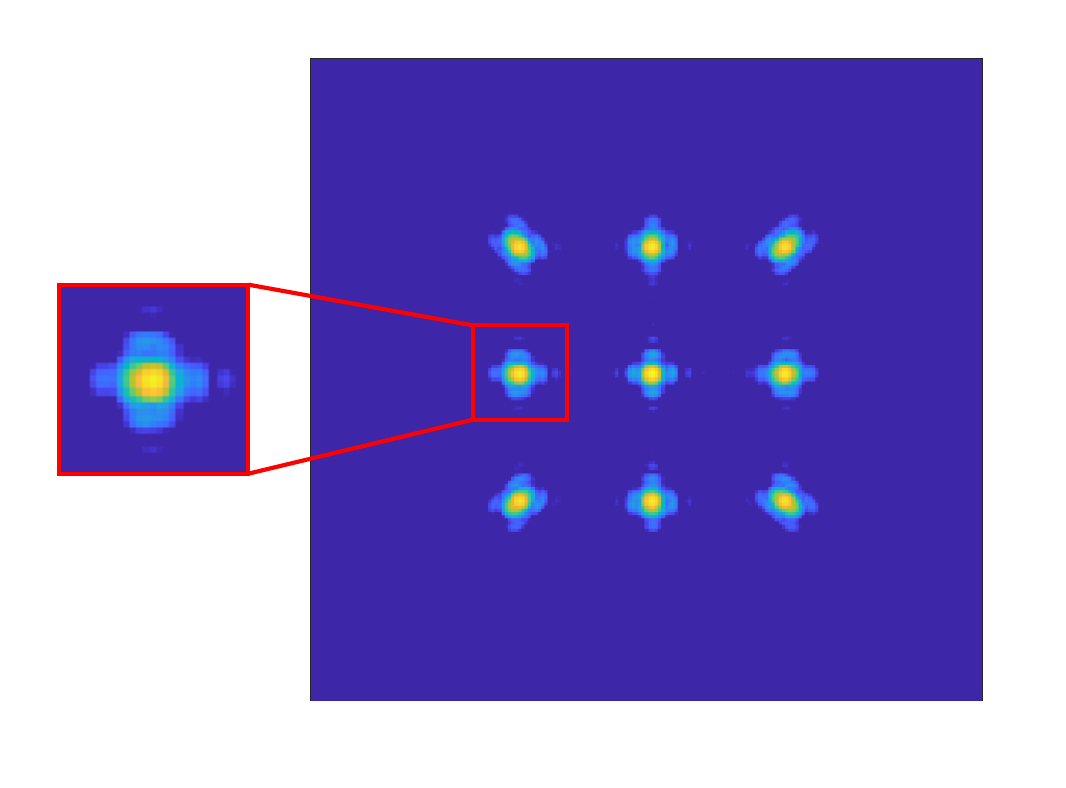}
\vspace{-0.8cm}
\caption{}
\label{fig-result-3-e}
\end{subfigure}
\begin{subfigure}[b]{0.4\linewidth}
\centering
\includegraphics[width=\linewidth]{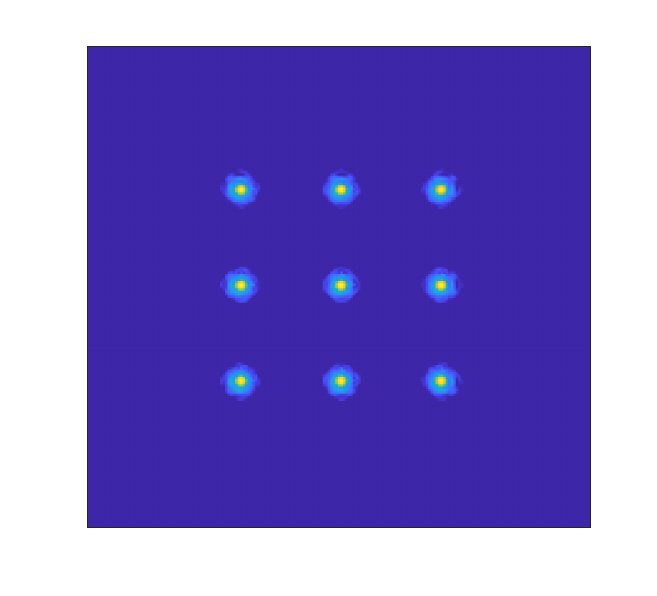}
\vspace{-0.8cm}
\caption{}
\label{fig-result-3-f}
\end{subfigure}

\caption{Imaging results of Algorithm \ref{ag1} with various RIS array sizes and phase shift tunable bits. RIS array size: (a), (c), and (e) for $100 \times 100$; (b), (d), and (f) for $200 \times 200$. RIS phase shift tunable bits: (a)-(b) for 1-bit; (c)-(d) for 2-bits; (e)-(f) for continuous. The horizontal and vertical directions of the images represent the y- and z-axis with the range of $[-25, 25]$, respectively. The color bars are the same as Figs. \ref{fig-result-1-b} and \ref{fig-result-1-c} with the range of $[-10{\rm{dB}},0]$.}
\label{fig-result-3}
\end{figure}

This subsection studies the influences of RIS array sizes and discrete RIS phase shifts on imaging performance.
The ROI is centered at the origin, and the side length is extended to $L^{\rm{i}} = 50$.
Nine point scatterers with unit scattering coefficients are located in the y-z plane with $x = 0$.
The measurement noise variance is 0.01.
The simulation results are displayed in Fig. \ref{fig-result-3}, where the 3D imaging results have been projected into the y-z plane for the convenience of performance analysis.
The left and right columns show the imaging results of the RIS arrays with $100 \times 100$ and $200 \times 200$ elements, respectively.
1-bit, 2-bit, and continuous RIS phase shifts are employed to derive the images in the first, second, and third rows, respectively.
RIS phase shifts are configured as the DFT matrix, and the nearest discrete phase shift values are chosen for Figs. \ref{fig-result-3-a}, \ref{fig-result-3-b}, \ref{fig-result-3-c}, and \ref{fig-result-3-d}.
Using the ECR recovery method given by \eqref{eq-recover-bt-fft} with IFFT techniques and the SAA, the point scatterers can be properly detected and imaged.
However, 1-bit discrete phase shifts may lead to additional artifacts at the corners of the images, as circled in Figs. \ref{fig-result-3-a} and \ref{fig-result-3-b}.
This is because the utilized RIS configuration matrix is a phase-discretized DFT matrix, but the original phase-continuous DFT matrix is employed to recover the ECR when using IFFT techniques, resulting in additional errors in ECR recovery.
The simulation results have shown that these errors can be effectively reduced by using higher discrete bits, and 2-bit phase shifts can achieve nearly the same results as continuous values, as depicted in Figs. \ref{fig-result-3-c} and \ref{fig-result-3-d}.
Moreover, the artifacts in Fig. \ref{fig-result-3-a} can be alleviated by increasing the RIS array size, as shown in Fig. \ref{fig-result-3-b}, owing to the richer information about the ROI captured by the larger RIS array size.
Additionally, the sidelobes are also reduced with large RIS arrays.
In conclusion, to efficiently obtain reasonable ROI images, at least 2-bit RIS phase shifts should be used, and the artifacts resulting from 1-bit RIS phase shifts can be alleviated by increasing the RIS array size.

Note that the ECR can be better recovered with the matrix inversion method given in \eqref{eq-recover-bt-matrix-inverse} even using discrete phase shifts, but the computational overhead may be greatly increased.
The execution times of the two ECR recovery methods are compared in Table \ref{tab-result-3}, where the simulation is conducted on a 2.6 GHz Intel Xeon Platinum 8358 Processor with 512 GB of RAM and MATLAB 2020a (64-bit).
ECR recovery with IFFT is significantly more efficient than matrix inversion.

\begin{table}[t]
  \renewcommand{\arraystretch}{1.5}
  \centering
  \fontsize{8}{8}\selectfont
  \captionsetup{font=small}
  \caption{Consumed times of different ECR recovery methods.}\label{tab-result-3}
  \begin{threeparttable}
    \begin{tabular}{c|c|c}
      \specialrule{1.2pt}{0pt}{-1pt}\xrowht{10pt}
      RIS array size & Matrix inversion \eqref{eq-recover-bt-matrix-inverse} & IFFT \eqref{eq-recover-bt-fft}\\
      \hline
      $100 \times 100$ & 2.9965s & 0.0002s\\
      \hline
      $200 \times 200$ & 365.54s & 0.0011s\\
      \specialrule{1.2pt}{0pt}{0pt}
    \end{tabular}
  \end{threeparttable}
\end{table}

\subsection{Comparison of Imaging Results with and without the RIS}
\label{sec-result-4-nlos}

In this subsection, simulations are conducted to validate the analysis presented in Sec. \ref{sec-case-study} and to demonstrate the advantages of using the RIS for imaging.
The ROI is defined as $\mathbb{D} = \{[x, y, z]^{\rm{T}} : -15 \le x \le 15, -15 \le y \le 15, -15 \le z \le 15\}$.
Four point targets are positioned within the ROI, with the simulation scenarios depicted in Fig. \ref{fig-nlos-scenario}.
It is assumed that both the RIS array and the virtual AP array comprise $100\times100$ antenna elements, spaced 1/2, and are oriented towards the ROI.
The reflection wall is positioned in the same plane as the RIS. The LOS paths between the ROI and the AP, as well as between the UE and the AP, are considered to be obstructed, simplifying the channel estimation process. Different received signal powers, with and without the RIS, necessitate employing path gain models from \cite{tang2021path} and \cite{goldsmith2005wireless} for the RIS-aided and wall-reflected paths, respectively. The transmit and receiving noise powers are set to -10 dBm and -50 dBm, respectively.

\begin{figure}
\centering
\captionsetup{font=footnotesize}
\begin{subfigure}[b]{0.325\linewidth}
\centering
\includegraphics[width=\linewidth]{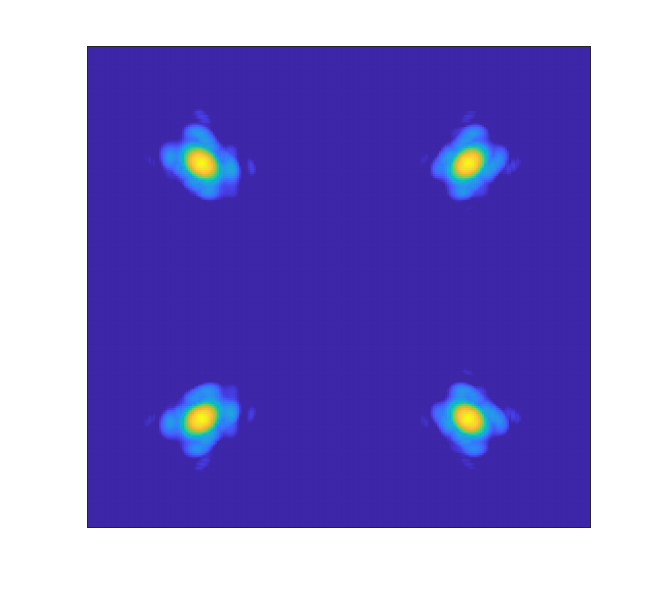}
\vspace{-0.8cm}
\caption{$D_{01}=50$}
\label{fig-result-with-ris}
\end{subfigure}
\begin{subfigure}[b]{0.325\linewidth}
\centering
\includegraphics[width=\linewidth]{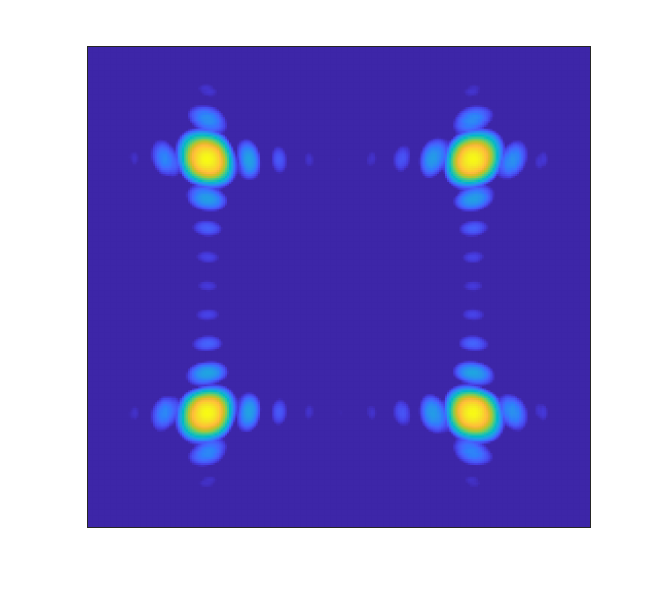}
\vspace{-0.8cm}
\caption{$D_{02}=82$}
\label{fig-result-without-ris-1}
\end{subfigure}
\begin{subfigure}[b]{0.325\linewidth}
\centering
\includegraphics[width=\linewidth]{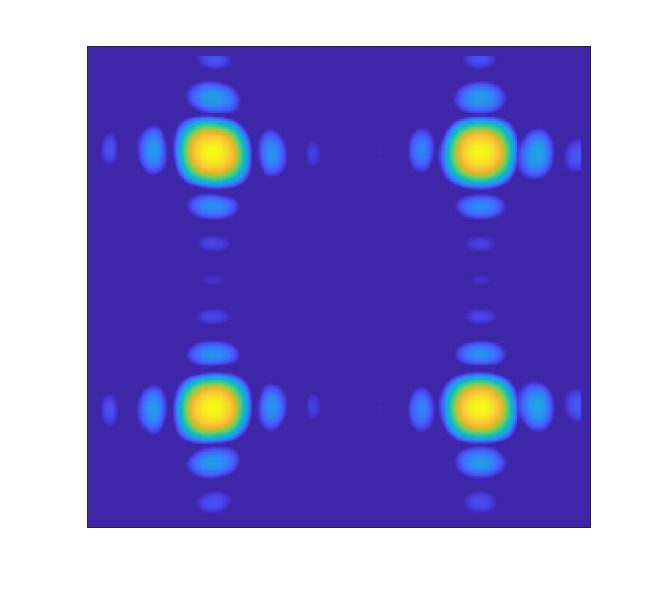}
\vspace{-0.8cm}
\caption{$D_{02}=121$}
\label{fig-result-without-ris-2}
\end{subfigure}

\caption{Imaging results of Algorithm \ref{ag1} with and without the RIS: (a) for with the RIS; (b) and (c) for without the RIS. The AP location is $[-20, 20, 20]^{\rm{T}}$ for (a) and (b), and $[20, 20, 20]^{\rm{T}}$ for (c). The horizontal and vertical directions of the images represent the y- and z-axis with the range of $[-15, 15]$, respectively. The color bars are the same as Figs. \ref{fig-result-1-b} and \ref{fig-result-1-c} with the range of $[-10{\rm{dB}},0]$.}
\label{fig-result-4-nlos}
\end{figure}

Simulation results, with an oversampling rate of four, are showcased in Fig. \ref{fig-result-4-nlos}.
With the RIS, the imaging distance $D_{01}=50$ enables high-resolution results and minimal sidelobes, as shown in \ref{fig-result-with-ris}.
Conversely, without the RIS, the imaging distance $D_{02}$ varies depending on the AP and reflector locations, with $D_{02}=82$ and $D_{02}=121$ for AP positions at $[-20, 20, 20]^{\rm{T}}$ and $[20, 20, 20]^{\rm{T}}$, respectively. This variation results in significantly lower imaging resolutions and higher sidelobes, with imaging performance deteriorating as $D_{02}$ increases, as depicted in Figs. \ref{fig-result-without-ris-1} and \ref{fig-result-without-ris-2}.
Although the signal power of the UE-ROI-RIS-AP path in Fig. \ref{fig-result-with-ris} is lower than that of the UE-ROI-reflector-AP paths in Figs. \ref{fig-result-without-ris-1} and \ref{fig-result-without-ris-2}, the proposed Algorithm \ref{ag1} efficiently suppresses additive noise and produces high-quality images. Hence, using the RIS for imaging significantly enhances imaging performance in the considered scenarios.

\subsection{Comparison of Various Imaging Algorithms with Limited Measurements}
\label{sec-result-5}

\begin{figure*}
\centering
\captionsetup{font=footnotesize}
\begin{subfigure}[b]{0.19\linewidth}
\centering
\includegraphics[width=\linewidth]{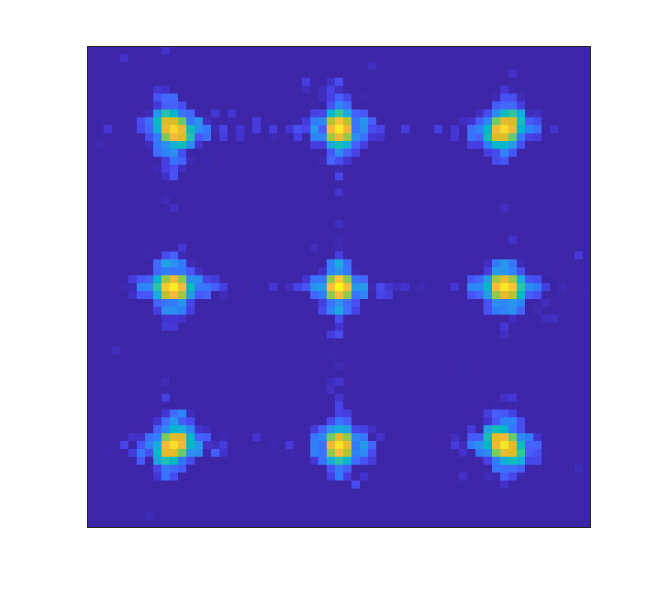}
\vspace{-0.85cm}
\caption{}
\label{fig-result-5-a}
\end{subfigure}
\begin{subfigure}[b]{0.19\linewidth}
\centering
\includegraphics[width=\linewidth]{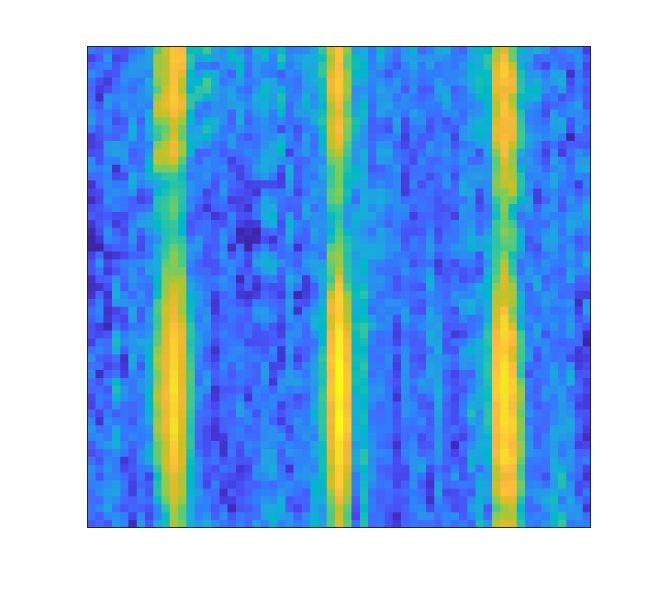}
\vspace{-0.85cm}
\caption{}
\label{fig-result-5-b}
\end{subfigure}
\begin{subfigure}[b]{0.19\linewidth}
\centering
\includegraphics[width=\linewidth]{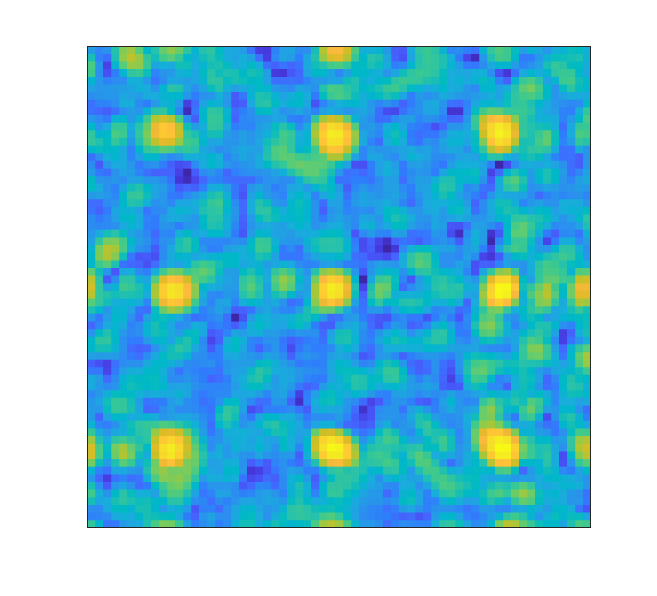}
\vspace{-0.85cm}
\caption{}
\label{fig-result-5-c}
\end{subfigure}
\begin{subfigure}[b]{0.19\linewidth}
\centering
\includegraphics[width=\linewidth]{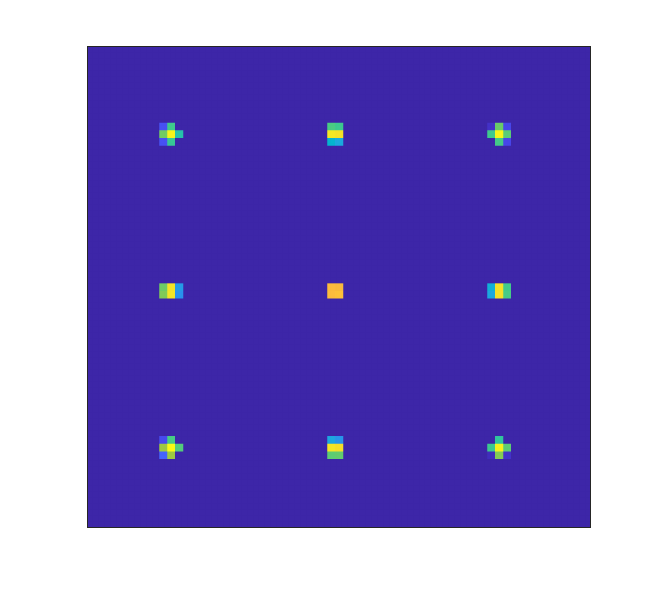}
\vspace{-0.85cm}
\caption{}
\label{fig-result-5-d}
\end{subfigure}
\begin{subfigure}[b]{0.19\linewidth}
\centering
\includegraphics[width=\linewidth]{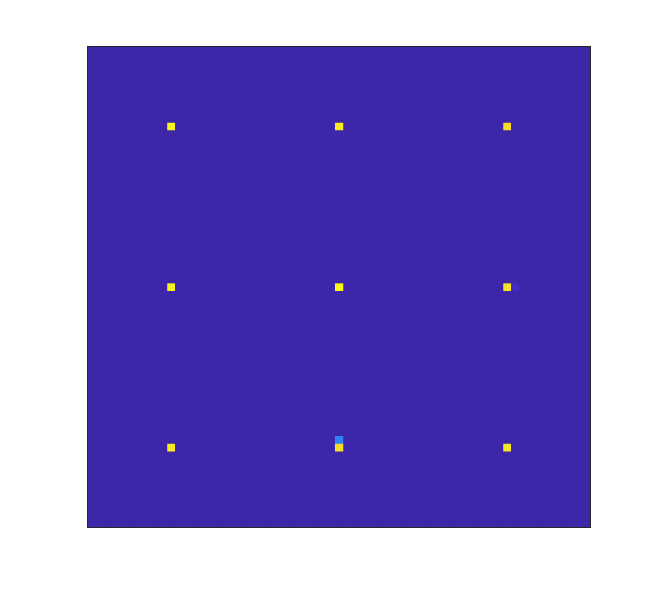}
\vspace{-0.85cm}
\caption{}
\label{fig-result-5-e}
\end{subfigure}

\begin{subfigure}[b]{0.19\linewidth}
\centering
\includegraphics[width=\linewidth]{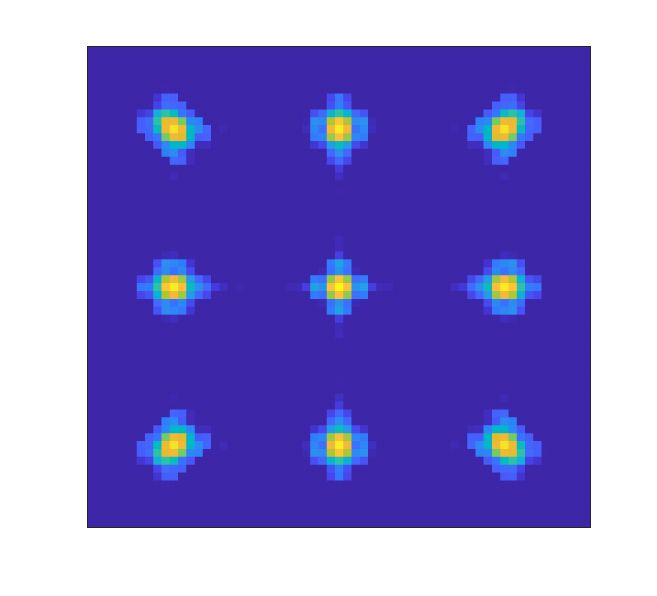}
\vspace{-0.85cm}
\caption{}
\label{fig-result-5-f}
\end{subfigure}
\begin{subfigure}[b]{0.19\linewidth}
\centering
\includegraphics[width=\linewidth]{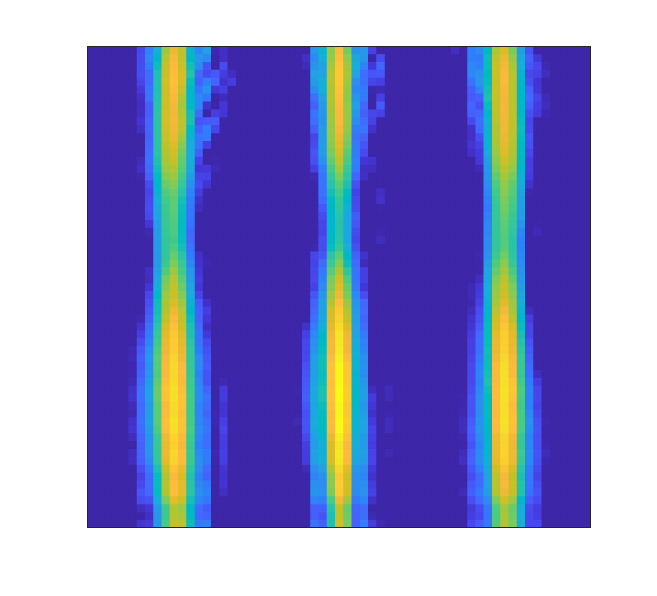}
\vspace{-0.85cm}
\caption{}
\label{fig-result-5-g}
\end{subfigure}
\begin{subfigure}[b]{0.19\linewidth}
\centering
\includegraphics[width=\linewidth]{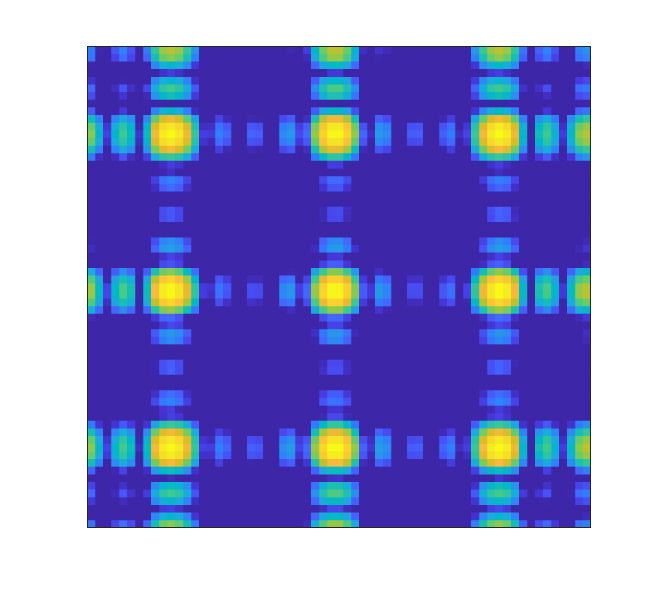}
\vspace{-0.85cm}
\caption{}
\label{fig-result-5-h}
\end{subfigure}
\begin{subfigure}[b]{0.19\linewidth}
\centering
\includegraphics[width=\linewidth]{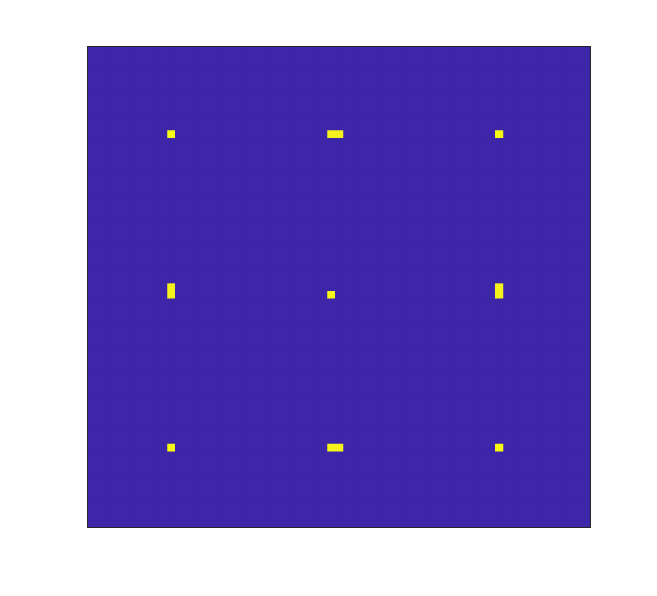}
\vspace{-0.85cm}
\caption{}
\label{fig-result-5-i}
\end{subfigure}
\begin{subfigure}[b]{0.19\linewidth}
\centering
\includegraphics[width=\linewidth]{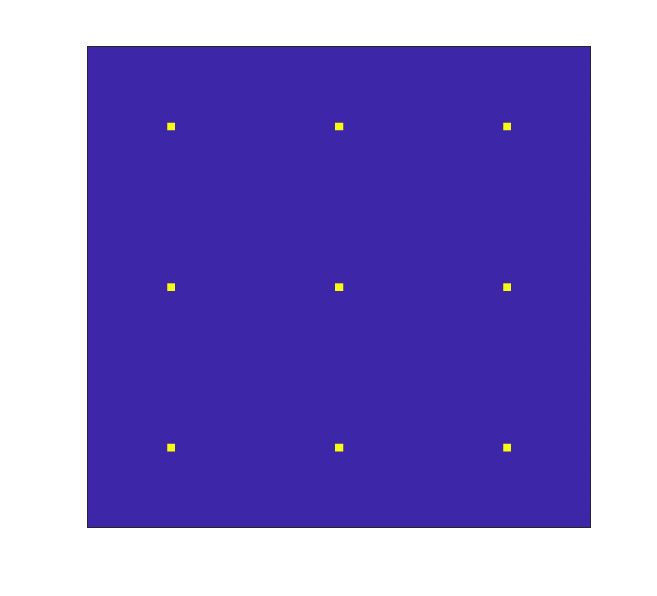}
\vspace{-0.85cm}
\caption{}
\label{fig-result-5-j}
\end{subfigure}

\begin{subfigure}[b]{0.19\linewidth}
\centering
\includegraphics[width=\linewidth]{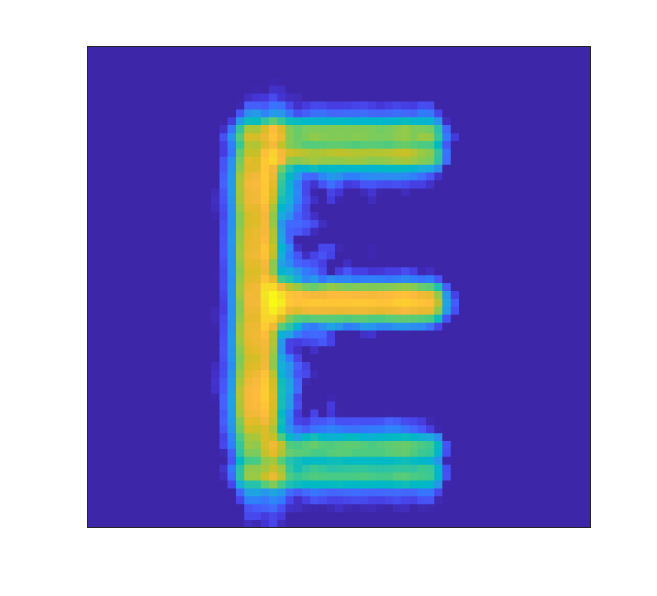}
\vspace{-0.85cm}
\caption{}
\label{fig-result-5-k}
\end{subfigure}
\begin{subfigure}[b]{0.19\linewidth}
\centering
\includegraphics[width=\linewidth]{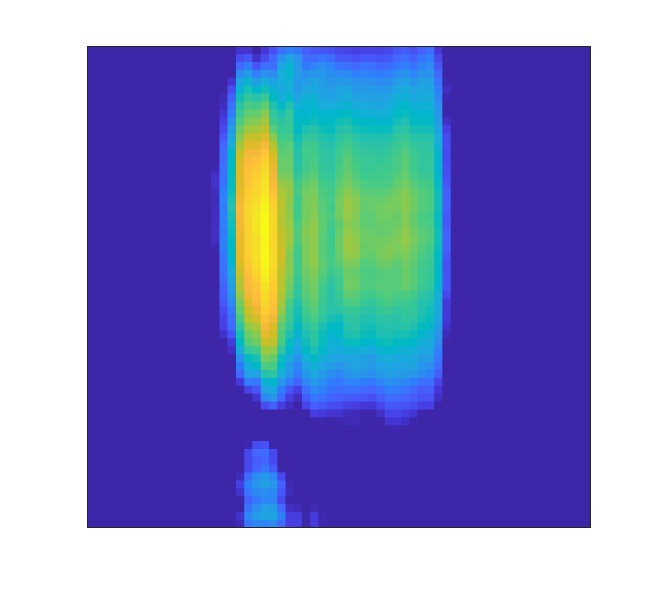}
\vspace{-0.85cm}
\caption{}
\label{fig-result-5-l}
\end{subfigure}
\begin{subfigure}[b]{0.19\linewidth}
\centering
\includegraphics[width=\linewidth]{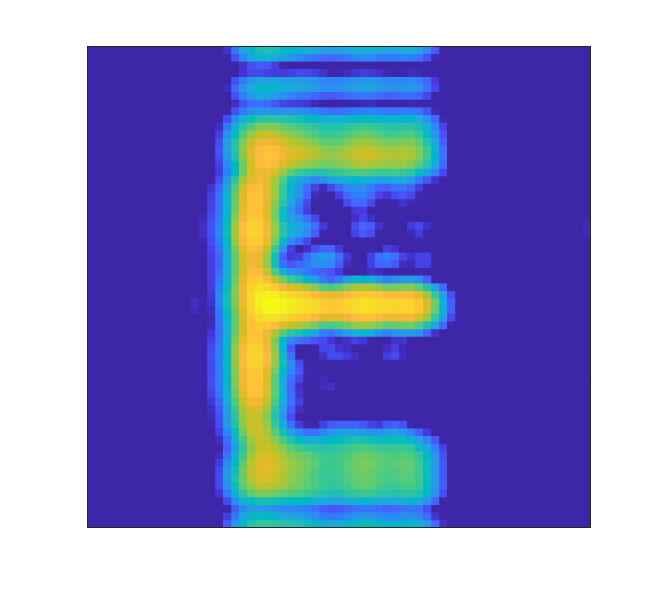}
\vspace{-0.85cm}
\caption{}
\label{fig-result-5-m}
\end{subfigure}
\begin{subfigure}[b]{0.19\linewidth}
\centering
\includegraphics[width=\linewidth]{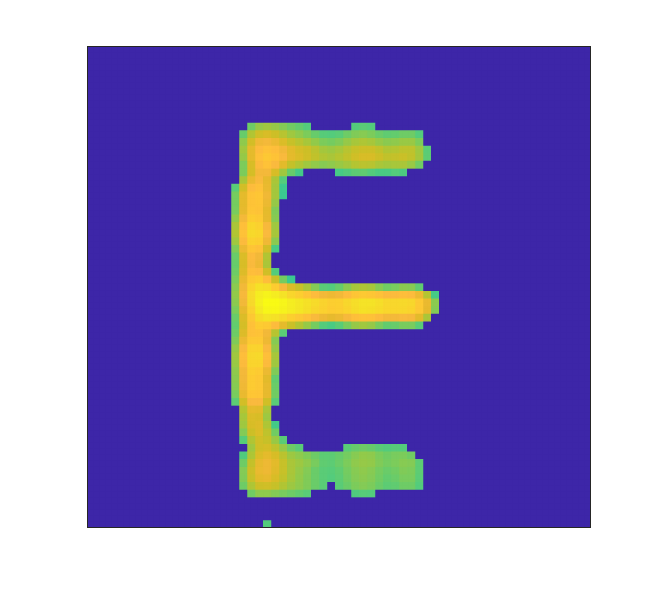}
\vspace{-0.85cm}
\caption{}
\label{fig-result-5-n}
\end{subfigure}
\begin{subfigure}[b]{0.19\linewidth}
\centering
\includegraphics[width=\linewidth]{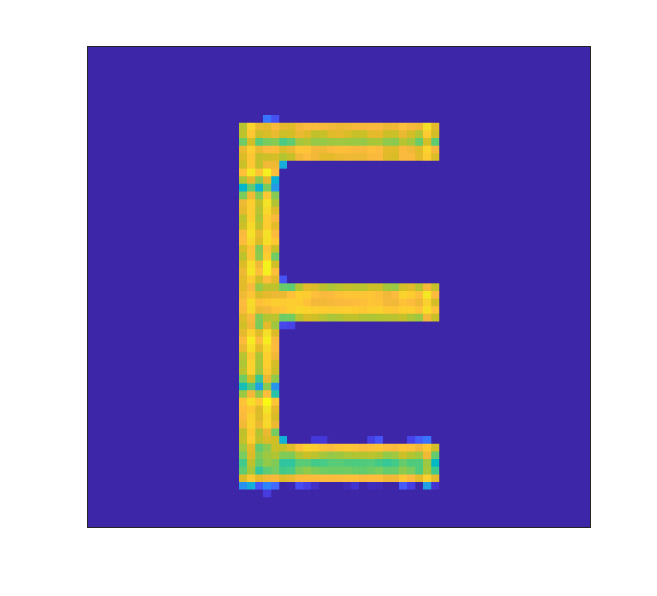}
\vspace{-0.85cm}
\caption{}
\label{fig-result-5-o}
\end{subfigure}

\caption{Imaging results of different algorithms with $R=M/16$. Imaging algorithms: (a), (f), and (k) for Algorithm \ref{ag1} with $R=M$; (b), (g), and (l) for the method of pseudo inverse matrix; (c), (h), and (m) for the method of block-controlled RIS; (d), (i), and (n) for the integrated algorithm of FT and CS in Algorithm \ref{ag3}; (e), (j), and (o) for the ISTA in Algorithm \ref{ag2}. Simulation scenarios: (a)-(e) for nine point targets with a noise variance of 10; (f)-(j) for nine point targets with a noise variance of 0.01; (k)-(o) for character ``E'' with a noise variance of 0.01. The horizontal and vertical directions of the images represent the y- and z-axis with the range of $[-15, 15]$, respectively. The color bars are the same as those in Figs. \ref{fig-result-1-b} and \ref{fig-result-1-c} with the range of $[-10{\rm{dB}},0]$.}
\label{fig-result-5}
\end{figure*}

In this subsection, we compare the performances of various imaging algorithms proposed in this study.
We consider the same ROI size as Sec. \ref{sec-result-4-nlos}, where nine point targets or the character ``E'' are located in the y-z plane with $x = 0$.
The character ``E'' involves numerous point scatterers.
We employ much fewer measurements than the Nyquist criterion with $R=M/16$.
The imaging results are exhibited in Fig. \ref{fig-result-5}, where the measurement noise variance is 0.01 or 10.
The voxel size is $1/2$, and zero-padding is used when employing the method of block-controlled RIS.
The simulations are executed on a desktop computer with a 2.5 GHz Intel(R) Core(TM) i7-11700 CPU, 32 GB of RAM, and MATLAB 2021b (64-bit).
The consumed time and memory are compared in Table \ref{tab-result-5}.

The first column of Fig. \ref{fig-result-5} presents the simulation results of Algorithm \ref{ag1} with $R=M$, serving as the benchmarks in this subsection, where the nine points and the shape of the character ``E'' are accurately recovered with unavoidable sidelobes.
Since Fig. \ref{fig-result-5-a} utilizes a large number of measurements, the background noise is slight compared to Fig. \ref{fig-result-5-f}.
The images formulated by the methods of pseudo-inverse matrix and block-controlled RIS are displayed in the second and third columns of Fig. \ref{fig-result-5}, respectively.
Figs. \ref{fig-result-5-b}, \ref{fig-result-5-g}, and \ref{fig-result-5-l} demonstrate that the method of pseudo-inverse matrix cannot realize imaging when $R=M/16$ because the differences between $\boldsymbol{\Omega}_R^{\rm{H}}\boldsymbol{\Omega}_R$ and $\mathbf{I}$ are high, and the recovered ECR from the UE to the RIS array is inaccurate.
In contrast, Figs. \ref{fig-result-5-c}, \ref{fig-result-5-h}, and \ref{fig-result-5-m} still derive rough shapes of the ROI with 6.25\% measurements of the Nyquist criterion but generate additional sidelobes and artifacts.
The two preliminary methods derive images with large background noise when $R=M/16$ with a noise variance of 10, as depicted in Figs. \ref{fig-result-5-b} and \ref{fig-result-5-c}.
Decreasing the noise variance can suppress the noise as shown in Figs. \ref{fig-result-5-g} and \ref{fig-result-5-h}.
Among the first three columns, the method of block-controlled RIS utilizes the least computational time since the dimensions of DFT and IDFT operations have been reduced by $1/16$, whereas the method of pseudo-inverse matrix occupies the most time since IFFT techniques are not used in ECR recovery.

\begin{table*}[t]
  \renewcommand{\arraystretch}{1.5}
  \centering
  \fontsize{8}{8}\selectfont
  \captionsetup{font=small}
  \caption{Consumed time of various imaging methods with various voxel sizes and the time and memory consumption of generating the sensing matrix (unless otherwise specified, $R = M/16$; unit for time: s).}\label{tab-result-5}
  \begin{threeparttable}
    \begin{tabular}{c|c|c|c|c!{\vrule width1.2pt}c|p{1.3cm}<{\centering}|p{1.3cm}<{\centering}}
      \specialrule{1.2pt}{0pt}{-1pt}\xrowht{10pt}
      \multirow{2}{*}{Voxel size} & \multirow{2}{*}{\makecell[c]{Algorithm \ref{ag1} \\ with $R=M$}} & \multirow{2}{*}{\makecell[c]{Method of pseudo \\ inverse matrix}} & \multirow{2}{*}{\makecell[c]{Method of block- \\ controlled RIS}} & \multirow{2}{*}{\makecell[c]{Algorithm \ref{ag3} \\ (per iteration)}} & \multirow{2}{*}{\makecell[c]{Algorithm \ref{ag2} \\ (per iteration)}} & \multicolumn{2}{c}{Sensing matrix generation} \\
      \cline{7-8}\specialrule{0pt}{0pt}{-1pt}
       & & & & & & time & memory \\\cline{7-8}
      \hline
      $1 / 4$ & 0.2866 & 0.3340 & {0.2147} & 1.3615 & /      & /      & /        \\
      $1 / 2$ & 0.1297 & 0.1890 & {0.0723} & 0.3755 & 1.9600 & 45.867 & 17.63 GB \\
      $1$     & /      & /      & {0.0265} & 0.1322 & 0.4847 & 11.078 & 4.269 GB \\
      \specialrule{1.2pt}{0pt}{0pt}
    \end{tabular}
  \end{threeparttable}
  \vspace{-0.1cm}
\end{table*}

Considering the method of block-controlled RIS leads to additional sidelobes and artifacts, which can be mitigated by exploiting the sparse nature of the ROI when $R<M$, we utilize this approach to demonstrate the effectiveness of the proposed Algorithm \ref{ag3}.
The maximum iteration number is set at $I_{\max} = 10$.
The imaging results, shown in the fourth column of Fig. \ref{fig-result-5}, indicate that the sidelobes resulting from FT-based algorithms are significantly reduced compared to those in the first three columns, achieving high imaging performance.
Notably, employing only 6.25\% of the measurements from the first column, Algorithm \ref{ag3} surpasses the imaging performance of Algorithm \ref{ag1} by leveraging the sparse property of the ROI. Additionally, background noise is further suppressed, even when the noise variance is increased to 10, as demonstrated in Fig. \ref{fig-result-5-d}, which represents the averaged outcome of 500 Monte Carlo simulations.

In the final analysis, although Algorithm \ref{ag2} delivers the superior imaging results within this subsection, it demands the most significant computational time and memory resources, mainly due to the generation of the sensing matrix and matrix-vector multiplications as indicated in \eqref{eq-ista-primary}.
The imaging results of Algorithm \ref{ag3} approximate those of Algorithm \ref{ag2} but involve additional sidelobes due to the unsatisfactory Nyquist criterion and the nature of FT-based imaging methods, i.e., $\boldsymbol{\epsilon}$ in \eqref{epsilon}.
However, Algorithm \ref{ag2} cannot achieve imaging with a voxel size of $1/4$, as the corresponding sensing matrix requires nearly 70 GB of memory, greatly exceeding the capacity of the desktop computer.
Furthermore, Algorithm \ref{ag2} utilizes a total of 65.467 seconds for the complete imaging process at a voxel size of 1/2, with 19.600 seconds dedicated to algorithm execution and 45.867 seconds to sensing matrix generation. In contrast, Algorithm \ref{ag3} requires only 3.755 seconds in total.
Therefore, Algorithm \ref{ag3} is capable of achieving imaging with smaller voxel sizes and significantly lower time and memory expenditures.

\subsection{Quantitative Performance Evaluation of Algorithm \ref{ag3}}
\label{sec-result-6}

\begin{figure}
    \centering
    \includegraphics[width=0.8\linewidth]{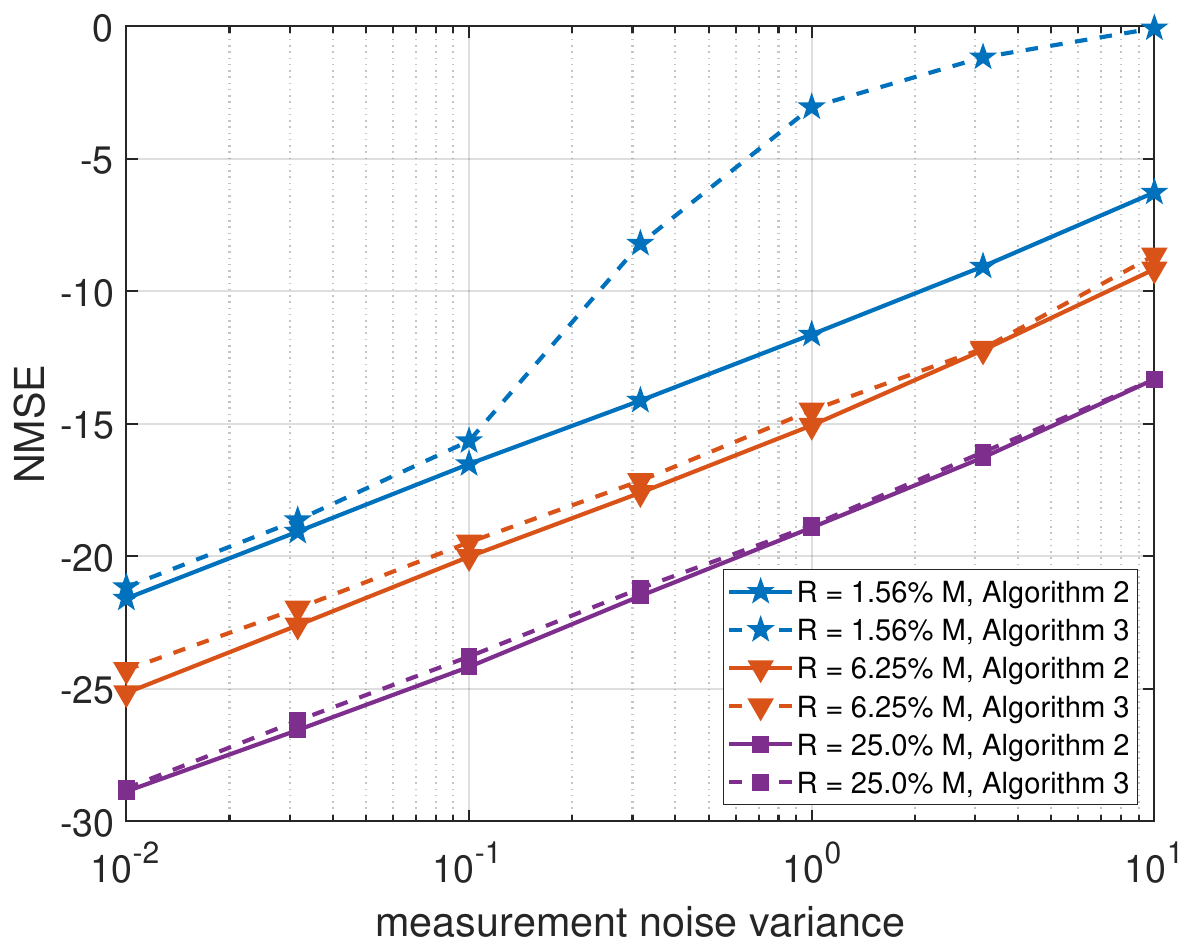}
    \captionsetup{font=footnotesize}
    \caption{NMSE performance with respect to the measurement noise variance and the number of measurements.}
    \label{fig-result-6-nmse}
\end{figure}

This subsection quantitatively assesses the imaging performance of Algorithm \ref{ag3} via the normalized mean square error (NMSE), defined as 
\begin{equation}
\mathrm{NMSE}=\frac{1}{I_{\mathrm{MC}}} \sum_{i=1}^{I_{\mathrm{MC}}} \frac{\left\|\widehat{\boldsymbol{\sigma}}^{(i)}-\boldsymbol{\sigma}\right\|_{2}^{2}}{\|\boldsymbol{\sigma}\|_{2}^{2}},
\end{equation}
where $\widehat{\boldsymbol{\sigma}}^{(i)}$ presents the estimated ROI image in the $i$-th simulation, and $\boldsymbol{\sigma}$ symbolizes the true ROI image.
The study includes $I_{\mathrm{MC}} = 1,000$ Monte Carlo simulations.
The simulation environment mirrors that described in Sec. \ref{sec-result-4-nlos}, featuring four point targets located in the y-z plane at $x=0$, each with a unit scattering coefficient. 
The relative subcarrier spacing is $\eta_{\Delta f} = 1/150$.
The method of block-controlled RIS is employed for imaging with a limited number of measurements, comparing Algorithms \ref{ag2} and \ref{ag3}.
The imaging outcomes of Algorithm \ref{ag2} are adjusted to normalize the highest scattering coefficient to one \cite{broquetas1998spherical}.
Fig. \ref{fig-result-6-nmse} displays the simulation results across various measurement noise variances and numbers of measurements, corresponding to different RIS element blocks. 

The simulations reveal that the integrated FT and CS imaging Algorithm \ref{ag3} can achieve approximately the same NMSE performance as the CS-based Algorithm \ref{ag2} when $R \ge 6.25\% M$, but with substantially lower time and memory demands, as previously discussed in Table \ref{tab-result-5}.
Nonetheless, as the number of measurements decreases to $R = 1.56 \% M$ and the measurement noise variance exceeds 0.1, Algorithm \ref{ag3} may produce images of inferior quality, diverging from  Algorithm \ref{ag2}.
This discrepancy stems from the substantial separation between RIS blocks ($4\lambda_0$ when $R = 1.56 \% M$), which greatly exceeds the Nyquist sampling criterion outlined in \eqref{eq-sampling-spatial}.
Consequently, this scenario incurs significant sidelobes and artifacts using the SAA algorithm. In such cases, elevated additive noise could mislead the thresholding operator detailed in \eqref{threshold}, causing sidelobes to be mistakenly identified as targets. In summary, Algorithm \ref{ag3} can achieve NMSE performance comparable to that of Algorithm \ref{ag2} under conditions of low noise variance or sufficient measurements.

\section{Conclusion}
\label{sec:conclusion}

In this study, we address the problem of FT-based radio imaging in RIS-aided communication systems. First, we propose a two-step wavenumber domain 3D imaging algorithm, where the ECR from the UE to the RIS array is recovered with specially designed RIS phase shifts, and the traditional SAA is subsequently adopted to generate the ROI image. Moreover, we derive the DRLs of the considered SIMO bistatic system, which are influenced by the system bandwidth, transmitting direction, working frequency, and RIS-subtended angle. Considering that communication systems typically have limited pilots, we propose two methods: pseudo-inverse matrix and block-controlled RIS to realize imaging with a small number of measurements. Furthermore, we make in-depth comparisons between FT- and CS-based imaging algorithms to reveal their different performances and application scenarios. Finally, we propose an integrated algorithm of FT and CS, where the large-dimensional sensing matrix is replaced with FT-based forward and backward operators, realizing high-quality imaging with low time and memory costs. Simulation results demonstrate that the proposed FT-based algorithms can realize real-time imaging, and the numerical resolutions approach the DRLs. Additionally, the integrated algorithm of FT and CS can achieve high-quality imaging with low measurement overheads, whereas the required time and memory resources are significantly reduced compared with CS-based algorithms.

\bibliographystyle{IEEEtran}
\bibliography{trans_ref}{}

\end{document}